\documentclass{article}
\pdfoutput=1
\usepackage{jheppub}

\usepackage{amssymb,amsmath,amsfonts,amsbsy,epsfig,wrapfig,caption,cancel,graphicx, pstricks, slashed, mathtools,xcolor}

\newcommand{\eq}[2]{\begin{equation}\label{#1}#2 \end{equation}}

\usepackage{xcolor}
\newcommand{\highlight}[1]{%
  \colorbox{blue!7}{$\displaystyle#1$}}

\newcommand{\de}{\delta}
\newcommand{\ep}{\epsilon}

\newcommand{\mpl}{M_{\rm pl}}
\newcommand{\mplf}{M_{\rm pl}}

\newcommand{\calL}{{\cal L}}
\newcommand{\calP}{{\cal P}}
\newcommand{\calR}{{\zeta}}

\newcommand{\calO}{{\cal O}}

\newcommand{\Q}{{\mathbf{q}}}

\newcommand{\Pb}{{\mathbf{p}}}
\newcommand{\K}{{\mathbf{k}}}
\newcommand{\Kp}{{\mathbf{k'}}}

\newcommand{\X}{{\mathbf{x}}}

\def\bea{\begin{eqnarray}}
\def\eea{\end{eqnarray}}
\def\be{\begin{equation}}
\def\ee{\end{equation}}
\def\ba{\begin{array}}
\def\ea{\end{array}}

\def\nn{\nonumber}

\definecolor{Dgreen}{rgb}{0,0.7,0.0}

\begin{document}
\author[a]{Adri\'an del Rio,}
\author[b]{Ruth Durrer}
\author[c]{and Subodh P.~Patil}
\affiliation[a]{Departamento de Fisica Teorica, IFIC. Centro Mixto Universidad de Valencia-CSIC.\\ Facultad de Fisica, Universidad de Valencia, Burjassot-46 100, Valencia, Spain.\\}
\affiliation[b]{ Dept. of Theoretical Physics, University of Geneva,\\ 24 Quai Ansermet, CH-1211 Geneva-4, Switzerland\\}
\affiliation[c]{Niels Bohr International Academy and Discovery Center,\\ Niels Bohr Institute, Blegdamsvej 17, Copenhagen, DK 2100, Denmark}

\date{\today}
\emailAdd{adrian.rio@uv.es, ruth.durrer@unige.ch, patil@nbi.ku.dk}
\title{\Large{Tensor Bounds on the Hidden Universe}}

\abstract{During single clock inflation, hidden fields (i.e. fields coupled to the inflaton only gravitationally) in their adiabatic vacua can ordinarily only affect observables through virtual effects. After renormalizing background quantities (fixed by observations at some pivot scale), all that remains are logarithmic runnings in correlation functions that are both Planck and slow roll suppressed. In this paper we show how a large number of hidden fields can partially compensate this suppression and generate a potentially observable running in the tensor two point function, consistently inferable courtesy of a large $N$ resummation. We detour to address certain subtleties regarding loop corrections during inflation, extending the analysis of \cite{SZ}. Our main result is that one can extract bounds on the hidden field content of the universe from bounds on violations of the consistency relation between the tensor spectral index and the tensor to scalar ratio, were primordial tensors ever detected. Such bounds are more competitive than the naive bound inferred from requiring inflation to occur below the strong coupling scale of gravity if deviations from the consistency relation can be bounded to within the sub-percent level. We discuss how one can meaningfully constrain the parameter space of various phenomenological scenarios and constructions that address naturalness with a large number of species (such as `N-naturalness') with CMB observations up to cosmic variance limits, and possibly future 21cm and gravitational wave observations.}

\maketitle
\pagebreak
\tableofcontents

\section{Introduction}

Observations strongly indicate that the Universe underwent an early phase of primordial inflation. Such an inflationary phase not only solves the horizon and flatness problems ~\cite{Guth,Linde:1981mu}, it also naturally produces a nearly scale invariant spectrum of density fluctuations~\cite{Mukhanov:1981xt} consistent with what has been observed in the cosmic microwave background (CMB). These fluctuations originated as quantum vacuum fluctuations that were forced out of the horizon by the quasi-exponential expansion of the Universe and subsequently squeezed, resulting in their phase coherence. The inflationary background also amplifies vacuum fluctuations of the transverse traceless part of the metric, leading to the generation of primordial gravitational waves~\cite{Starobinsky:1979ty} as well as fluctuations of all other fields present in the quantum vacuum whether they couple directly to the inflaton or not.

In this paper we consider the effects of fields that one would ordinarily be tempted to ignore during inflation: hidden fields, defined as fields that couple only to gravity and have no direct couplings to the inflaton. In their adiabatic vcauum, such fields would only serve to renormalize background quantities\footnote{Whose effects therefore would simply be absorbed into physical measurements of quantities such as $\epsilon := -\dot H/H^2$ (e.g. through the detection of primordial tensors) and its derivatives or the ratio $H^2/\mpl^2$, all of which denote renormalized quantities.} and induce unobservably small (i.e. Planck and slow roll suppressed) logarithmic runnings in cosmological correlation functions. However, in large enough numbers, their effects can add up to an observable running of the spectral index of the two point function of the tensor perturbation, consistently inferable via a "large $N$" expansion that allows us to resum a restricted class of diagrams. The running induced for correlation functions of the curvature perturbation on the other hand remains feeble, since the relative suppression of the interaction vertices by factors of $\epsilon$ is too great to be overcome by large $N$ and still consistent with being below the strong coupling scale of gravity.   

One can thus use this observation to convert bounds on the violation of the tensor to scalar consistency relation to a bound on the possible number of hidden fields present in the universe with masses below the scale of inflation, \textit{were primordial tensors ever to be observed}\footnote{Although fields with masses much greater than the Hubble scale during inflation also contribute to the running of the tensor spectrum, their effects are very suppressed at long wavelengths and so will not contribute to the bounds derived here. Fields with masses $m \sim H$ (cf. \cite{Chen:2009we,Chen:2009zp}) do not affect the running of two point functions, although they can imprint on higher order (cross-)correlation functions with additional interactions not considered here \cite{Saito:2018omt}.}. For simplicity, we focus on hidden scalars, although our argument generalizes straightforwardly to particles of other spin \cite{DDP}. We find that any bound from above (to some confidence level) on deviations from the tensor to scalar consistency relation  
\eq{}{n_t +  \frac{r_*}{8} \lesssim \xi} 
for some positive $\xi$, translates into a bound on the number of hidden species as
\eq{final0}{N\lesssim 8.5\times 10^2 \frac{\xi}{r_*^2} \Delta_\zeta^{-1}}
Where $\Delta_\zeta \approx 2.44 \times 10^{-9}$ \cite{Aghanim:2015xee} is the amplitude of the spectrum of the curvature perturbation at the pivot scale where we determine the tensor to scalar ratio $r_*$, with $n_t$ being the tilt of the tensor spectrum. If we presume the most optimistic case that $r_* \sim 0.06$ then the best we can hope to bound $N$ through CMB measurements is by
\eq{}{N \lesssim \frac{3.5 \times 10^{11}}{r_*^2}\xi \sim 10^{14}\times \xi}
We note that this bound is only interesting if it is stronger than the bound coming from the requirement that we stay below the scale at which gravity becomes strongly coupled \cite{Dvali:2007hz, Dvali:2007wp} (cf. eq (\ref{lqg}), reviewed in appendix \ref{a:sc}):
\eq{sc0}{N \lesssim \frac{16\pi^2\mpl^2}{H^2}= \frac{32 \Delta_\zeta^{-1}}{r_*} \approx \frac{1.3\times10^{10}}{r_*} \sim 10^{11}.}
In order to infer a stronger bound from (\ref{final0}) than from consistency imposed by being below the strong coupling scale (\ref{sc0}), we would need to bound $\xi$ one order of magnitude better than we the accuracy with which we measure $r_*$. As we shall elaborate upon further, cosmic variance limits us to bounds on $\xi$ no better than the percent level (were $r \sim 0.06$) from CMB observations alone, allowing only marginally to  bound the parameter space of a variety of models that attempt to address the hierarchy problem with a large number of sectors \cite{Dvali:2009ne, Arkani-Hamed:2016rle}. However, as we discuss further, observations of the stochastic background at very different comoving scales through future 21cm and space based gravitational wave interferometer observations could allow us to entertain significant improvements upon these constraints.
 
We begin this paper with an outline of our calculation with details deferred to the appendix. It behoves us to elaborate upon various subtleties encountered in the calculation of loop corrections to cosmological correlation functions relevant to this calculation \cite{SZ,Adshead}. In particular, we extend the analysis of Senatore and Zaldarriaga \cite{SZ} which pointed out that dimensional regularization had only been partially implemented in previous calculations (e.g. \cite{weinberg} and subsequent studies), where it was found that loop corrections induced a running of the form $\log (k/\mu)$ in the two point function of the curvature perturbation, with $\mu$ some arbitrary renormalization scale. Including previously neglected corrections to the mode functions and to the integration measure in $D = 3 + \delta$ spatial dimensions\footnote{A conclusion independently arrived at by working in a \textit{mass dependent} regularization scheme (a hard cutoff in physical momenta).}, it was found that loops instead induce a correction of the form $\log (H/\mu)$ \cite{SZ}. 

At first glance this appears to preclude any running of the loop correction, which cannot be the case in general as quantum corrections typically induce scale dependence unless we are at a fixed point of the theory, e.g. in the dS (de Sitter) limit where an exact dilatation (i.e. scale) invariance is realized -- implicitly assumed in \cite{SZ}. Since corrections to the correlation functions are being forged as modes exit the horizon during single clock inflation, it must be the case that what appears inside the log is in fact $H_k$ -- the Hubble scale at the time the $k$-mode exits the horizon. We demonstrate this explicitly in appendix \ref{a:tensor}, where we show how additional slow roll corrections to the mode functions and the integration measures within the loop integrals indeed result in a correction of the form $\log (H_k/\mu)$. Upon fixing the renormalization conditions at some (pivot) scale $\mu = H_*$, one reintroduces a running as one moves away from this scale, but now of the form $\log (H_k/H_*) \to -\epsilon\, \log(k/k_*)$. This contribution to the running is far too feeble to ever be observed for the curvature perturbation\footnote{In section \ref{s:disc}, we discuss the possible implications of the running of the two point function of the curvature perturbation for whether or not a given model of inflation is eternal according to criteria derived in \cite{eternal}.}, but does have a potentially observable effect on the tilt of the tensor spectrum. In Section~\ref{s3} we derive the modifications of the tensor and scalar spectral indices due to the presence of hidden fields and in Section~\ref{s:disc}, we discuss possible observational bounds on $N$ and generalizations of our results.
\\

\noindent {\bf Notation:} In what follows, we shall consider a spatially flat FRLW universe with line element in Cartesian coordinates 
\be ds^2 = a^2(\tau)\left[-d\tau^2 +\de_{ij}dx^idx^j\right] =  g_{\mu\nu}dx^\mu dx^\nu \,, \label{e:Fried}\ee 
where $\tau$ denotes conformal time and physical time is given by $dt=ad\tau$. Derivatives w.r.t. $\tau$ are denoted by a prime and those w.r.t. $t$ by an overdot. The physical Hubble parameter is $H=\dot a/a$. 

\section{Outline of the calculation}
We consider an inflationary Universe with an inflaton $\phi$ taken to be the only field with an evolving background (hence energy density) and $N$ additional hidden scalar fields $\chi_n$ with a flat target space, minimally coupled to gravity and taken to be in their respective adiabatic vacuum states. We only consider hidden fields with masses $m^2\ll H^2$, which can therefore be treated as effectively massless but are quantum mechanically excited by the background expansion\footnote{Equivalently, thermally excited with the identification $T_{\rm dS} = \frac{H}{2\pi}$ in units where $k_B =\hbar=c= 1$.} during inflation. By assumption the $\chi_n$ have no non-gravitational interactions. The action is then given by
 \begin{eqnarray}
\label{e2:act}S &=& \frac{\mplf^2}{2}\int d^4x\sqrt{-g}R[g] - \frac{1}{2}\int d^4x\sqrt{-g}\left[\partial_\mu\phi\partial^\mu\phi + 2V(\phi)  
+\sum_{n=1}^{N}\partial_\mu\chi_n\partial^\mu\chi_n\right]  \,,\end{eqnarray}
where $\mplf =(8\pi G)^{-1/2}$ is the reduced Planck mass\footnote{This so far bare quantity also gets renormalized via diagrams involving external graviton legs with loops of massive fields. However, in the massless limit, the contributions of each species to the divergent and finite parts of $\mpl^2$ and the cosmological constant vanishes \cite{BD} whilst still lowering the strong coupling scale (cf. appendix \ref{a:sc}).}. We presume the background to be quasi de-Sitter, such that
\be
\ep := \frac{\dot\phi^2_0}{2H^2\mplf^2}  =-\frac{\dot H}{H^2}\ll 1 \, \label{slowrollp}\,,
\ee
so that $H^2 = V(\phi_0)/(3\mplf^2)\sim$ const, and for completeness we define higher order slow roll parameters $\ep_i$ as
\be
 \ep_1\equiv \ep\,, \qquad \ep_{i+1}=\frac{\dot \ep_i}{H\ep_i}\,, \quad i\geq 1 \,. \label{slowrolldefinitions}
\ee
In order to discuss perturbations around this background, we first ADM decompose the metric as
\eq{adm0}{ds^2 = -N^2dt^2 + h_{ij}(dx^i + N^idt)(dx^j + N^jdt),}
and work in comoving gauge, defined to be the foliation in which we have gauged away the inflaton fluctuations. In this gauge, the only dynamical degrees of freedom are contained in the 3-metric $h_{ij}$ which has now acquired, or `eaten' a scalar polarization that was the inflaton fluctuation \cite{Cheung:2007st}\footnote{We note that comoving gauge is defined by the vanishing of $\delta T^0_i$. This is still satisfied in the presence of an arbitrary number of hidden fields since their contributions to $\delta T^0_i$ go as $\dot\chi\partial_i\chi$ which vanishes identically since by assumption the $\chi$ fields have no classically evolving background. Note that this statement persists at the quantum level as well, since $\langle\dot\chi\partial_i\chi\rangle = 0$ by isotropy of the Bunch-Davies vacuum state.}     
\bea
\label{cgf0}
\phi(t,x) &=& \phi_0(t),\\
h_{ij}(t,x) &=& a^2(t)e^{2\calR(t,x)}\hat h_{ij},~~~ \hat h_{ij} = {\rm exp}\left[\gamma_{ij}\right],
\eea
where $\gamma^i_i = \partial_i\gamma^i_j = 0$ is (transverse traceless) graviton, and $\zeta$ is the comoving curvature perturbation. The quasi dS background then results in a nearly scale invariant spectrum of curvature perturbations ~\cite{mukbook,ruthbook,weinbergbook}
\be\label{e2:inf-spec}
\calP_\zeta(k) = \frac{H_*^2}{8\pi^2\ep \mplf^2}\left(\frac{k}{k_*}\right)^{n_s-1} \, ,\qquad
n_s-1= -2\ep-\ep_2\,.
\ee
In addition helicity 2 tensor perturbations of the metric are amplified from their initial quantum vacuum state leading to a power spectrum for primordial gravitational waves given by
\be\label{e2:inf-spec-tensor}
\calP_\gamma(k) = \frac{2H_*^2}{\pi^2 \mplf^2}\left(\frac{k}{k_*}\right)^{n_t} \, ,\qquad
n_t= -2\ep \,.
\ee
The ratio of these two quantities 
\be
\label{tsr}
r = \frac{\calP_\gamma(k_{*})}{\calP_\zeta(k_{*})} =16\ep
\ee
defines the tensor to scalar ratio. Note that its value, as well as $n_t$ and $n_s$ depend on the pivot scale, $k_*$, and $H_*$ is defined as the value of the Hubble parameter at the time the mode $k_*$ exits the horizon. The single field scalar tensor consistency relation is simply the identity $r+8n_t=0$. At present, no tensor perturbations have been identified in the observed CMB anisotropies and an upper limit of $r\lesssim 0.06$ has been derived for $k_*=0.002$ Mpc$^{-1}$ \cite{planck, bicep}. We remind the reader that there are higher order corrections to the tilt of the scalar and tensor spectra that come from the background dynamics alone, which we will return to later. We are interested in additional corrections to these from virtual effects due to the presence of the hidden fiends $\chi_n$.  

\subsection{Diagrammatic preliminaries in the `in-in' formalism}
Perturbing the action (\ref{e2:act}) in comoving gauge (\ref{cgf0}) results in the quadratic action
\eq{s2r0}{S_{2,\calR} = \mplf^2\int d^4x\, a^3\,\epsilon\left[\dot\calR^2 - \frac{1}{a^2}(\partial\calR)^2 \right]}
\eq{s2h0}{S_{2,\chi} = \frac{1}{2}\int d^4x\, a^3\left[\dot\chi_n{\dot\chi_n} - \frac{1}{a^2}\partial_i\chi_n\partial_i\chi_n\right]}
\eq{s2t0}{S_{2,\gamma} = \frac{\mpl^2}{8}\int d^4x\, a^3\left[\dot\gamma_{ij}\dot\gamma_{ij} - \frac{1}{a^2}\partial_k\gamma_{ij}\partial_k\gamma_{ij}\right]}
and the cubic interaction vertices
\eq{cubint0}{S_{3, \calR \chi} = \int d^4x\,a^3  \epsilon\left[\frac{\calR}{2}\left(\dot\chi_n{\dot\chi_n} + \frac{1}{a^2}\partial_i\chi_n\partial_i\chi_n\right) - \dot\chi_n\partial_i\chi_n\partial_i\partial^{-2}\dot\calR\right] }
and
\eq{cubintT}{S_{3, \gamma \chi} = \frac{1}{2}\int d^4x\,a\left[\gamma_{ij}\partial_i\chi_n\partial_j\chi_n \right] = \frac{1}{2}\int d^4x\,a\gamma_{ij}\Pi^\chi_{ij}.}
Where $\Pi^\chi_{ij}$ is the anisotropic stress of the $\chi$ fields and the sum over $n$ is implicit. The form of (\ref{cubint0}) -- in particular its $\epsilon$ suppression -- is not immediately obvious from naively expanding the original action (\ref{e2:act}) having solved for the lapse and shift constraints, which results in an expression that is nominally unsuppressed in $\epsilon$ (\ref{s3h}). However as shown in appendix \ref{a:vert}, similar to what occurs for the cubic and higher order self interactions for $\zeta$ \cite{Maldacena}, enough integrations by parts show that the $\zeta\chi\chi$ cubic (and the $\zeta\gamma\chi\chi$ quartic) interactions are suppressed by an overall factor of $\epsilon$. Similarly, interactions that are higher order in $\zeta$ will be sequentially suppressed by additional powers of $\epsilon$, consistent with its nature as an order parameter parameterizing the breaking of time translational invariance by slow roll \cite{Chluba}.

We are interested in calculating the finite time correlation functions of the curvature perturbations
\eq{psdef}{\frac{k^3}{2\pi^2} \langle\calR_{\K}(\tau)\calR_{\Q}(\tau)\rangle := (2\pi)^3\delta^3(\K + \Q)\calP_\zeta(k),}
and tensor perturbations $\gamma^r_{ij}$
\eq{pstdef}{\frac{k^3}{2\pi^2} \langle\gamma^r_{ij,\K}(\tau)\gamma^r_{ij,\Q}(\tau)\rangle := (2\pi)^3\delta^3(\K + \Q)\calP_\gamma(k),}
where we have summed over the two independent polarizations. Both of the above are of the form $\langle\mathcal O(\tau)\rangle$ where the angled brackets denote expectation values with a given initial density matrix (which we take to correspond to the Bunch-Davies vacuum), unitarily evolved forward in the interaction picture with the Dyson operator
\eq{dop}{U(\tau,-\infty) = T\,{\rm exp}\left(-i\int^\tau_{-\infty}H_I(\tau') d\tau' \right),}
where $T$ denotes time ordering and where $H_I$ is the interaction Hamiltonian (equal to minus the interaction Lagrangian given in eqs. (\ref{cubint0}) and  (\ref{cubintT}) respectively for the interactions in question \cite{weinberg}. Reading right to left, one evidently evolves the Bunch-Davies vacuum from the initial time $-\infty$ to $\tau$, inserts the corresponding free-field operator $\mathcal O_0(\tau)$ at time $\tau$ and then evolves back to $-\infty$:
\eq{iidef0}{\langle\calO(\tau)\rangle = {\langle 0_{\rm in}|\left[T\,{\rm exp}\left(-i\int^\tau_{-\infty} H_I(\tau') d\tau' \right)\right]^\dagger \calO_0(\tau)\left[T\,{\rm exp}\left(-i\int^\tau_{-\infty} H_I(\tau') d\tau' \right)\right]|0_{\rm in}\rangle}}
The above can be shown to be formally equivalent to the expression \cite{weinberg}
\eq{inin}{\langle\calO(\tau)\rangle = \sum_{n=0}^\infty i^n \int^\tau_{-\infty}d\tau_n\int^{\tau_n}_{-\infty}d\tau_{n-1}...\int^{\tau_2}_{-\infty}d\tau_1\langle[H_I(\tau_1),[H_I(\tau_2),...[H_I(\tau_n),\calO_0(\tau)]...]] \rangle}
provided one is mindful of how one selects the correct initial interacting vacuum \cite{Adshead} -- an important point that we will return to shortly. 
\begin{figure}[t!]
	\hfill
	\begin{minipage}[t]{0.49\textwidth}
		\begin{center}
			\vspace{28.3pt}
			\epsfig{file=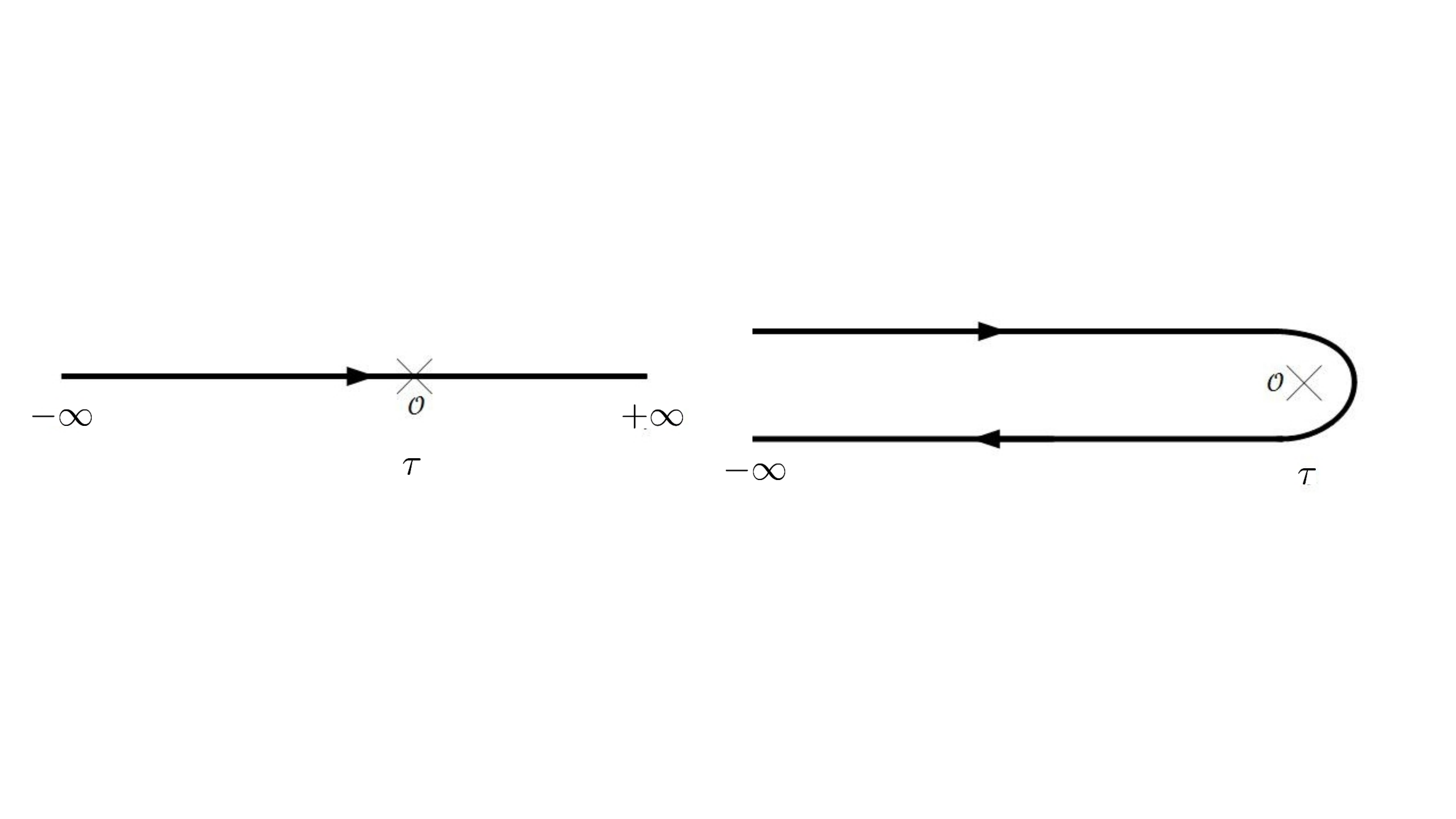, height=1.8in, width=3.1in}
			\caption{The S-matrix contour (left) compared to the Schwinger-Keldysh contour (right).\label{f:inin}}
		\end{center}
	\end{minipage}
	\hfill
	\begin{minipage}[t]{0.49\textwidth}
		\begin{flushleft}
			\epsfig{file=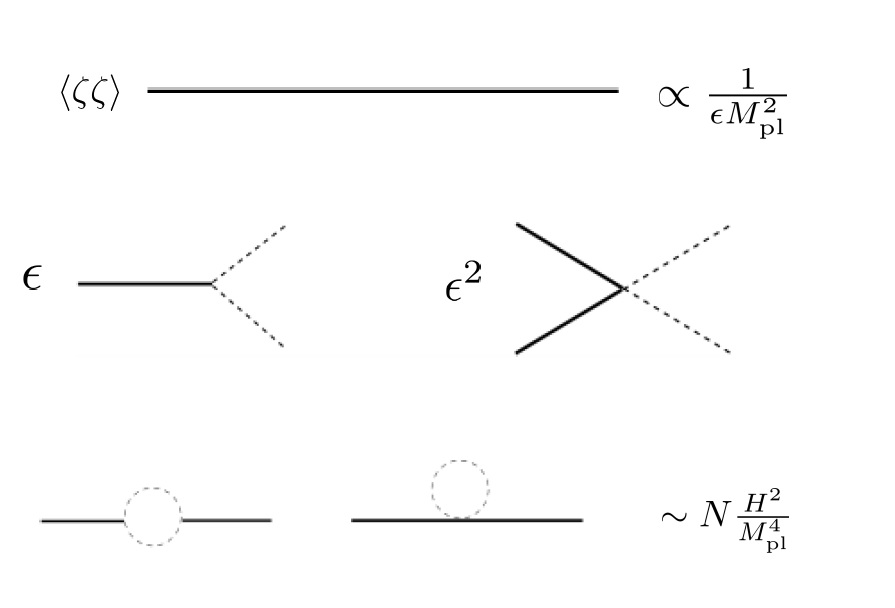, height=2.2in, width=3.1in}
			\caption{One loop corrections to $\langle\zeta\zeta\rangle$. Solid lines denote the curvature perturbation propagator, dashed lines denote the $\chi$-propagator.
			\label{f:1loop}}
		\end{flushleft}
	\end{minipage}
	\hfill
\end{figure}

Although useful for practical purposes, such an expectation value does not lend itself to the usual diagrammatic expansion one uses when dealing with S-matrix elements. In order to implement this one can equivalently consider the expression (\ref{iidef0}) as the product of an arbitrary operator $\calO_0(\tau)$ with the unitary evolution operator:
\eq{iidefc}{\langle\calO(\tau)\rangle = {\langle 0_{\rm in}| T_C\,\left[{\rm exp}\left(-i\oint H_I(\tau') d\tau' \right)\calO_0(\tau)\right]|0_{\rm in}\rangle}}
with the contour going from $-\infty \to \tau$ and back again (cf. Fig. \ref{f:inin}), and with $T_C$ denoting contour ordering with fields living on the reverse contour treated as independent fields for intermediate manipulations, only being set equal to the original fields at the end of the calculation. Due to its formal similarity with an S-matrix element, the former does indeed lend itself to a diagrammatic expansion which we will not make explicit use of in the following, but we nevertheless find useful for reasoning diagrammatically.

Suppressing the difference between the fields that live on the future and past directed contours (as a result of which there are typically many cancellations as one sums up relevant diagrams) as  shorthand, one can nevertheless intuit the parametric and external momentum dependences of the various graphs that one can write down. For example, at one loop, one has two possible contributions to the correction to the two point correlation function of the curvature perturbation\footnote{There are also contributions from cubic interactions involving $\zeta$ alone, but these will be suppressed by two extra powers of $\epsilon$ \cite{Maldacena, Sloth}.} as indicated in Fig. \ref{f:1loop}. However only the diagram involving two cubic vertices results in a dependence on the external momenta\footnote{Although vanishing for massless fields, the quartic `seagull' interactions contributes to wavefunction renormalization for any small but finite mass, accounted for in practice by fixing the (fully renormalized) expressions $H_*^2/\mpl^2$ and $\epsilon_*$ via the amplitude of the power spectrum and the tensor to scalar ratio at some pivot scale $k_*$.} and hence contributions to the running of the spectral index, which is the object of our interest.

At two loops, we notice that the double sunset graphs (involving two independent loops of hidden fields) dominate when $N \gg 1/\epsilon$ relative to all other contributions (Fig. \ref{f:2loops})\footnote{There are an additional two loop diagrams corresponding to a single sunset graph with a tadpole insertion to an internal $\chi$ propagator, but this is accounted for by wavefunction renormalization of the $\chi$ fields and considering diagrams with internal lines taken to be renormalized propagators when summing graphs.}. This structure persists at each loop order and permits the resummation of a restricted subset of diagrams (consisting only of the sunset diagrams) in the large $N$ limit, allowing us to consistently infer the running even in the event that it could compete with the running induced from the background dynamics alone. 

It is here that we lose interest in the corrections to the running of the curvature perturbation, since it will turn out that no amount of enhancement by factors of $N$ can overcome the slow-roll suppression of the corrections, consistent with the strong coupling bound (\ref{sc0}). This is in part because of the $\epsilon$ suppression of the interaction vertices (\ref{cubint}) and (\ref{quint}), but also since (as we shall see shortly) the corrections must be of the form $\log H_k/\mu$, as opposed to the $\log k/\mu$ which eventually introduces additional slow-roll suppression. Tensor perturbations on the other hand, have interactions that are unsuppressed by $\epsilon$ and will have potentially observable consequences, which we turn to presently.  

\subsection{Running of the tensor two-point function}

For the rest of this paper, we shall be interested in the operator expectation value (\ref{pstdef}), which is shorthand for
\eq{inin1}{\langle \gamma^s_{ij, \K}(\tau)\gamma^{s'}_{ij,\Kp}(\tau) \rangle = \langle \left(T\,e^{-i\int^\tau_{-\infty } d\tau'H_I(\tau')} \right)^\dag \gamma^{0,s}_{ij, \K}(\tau)\gamma^{0,s'}_{ij,\Kp}(\tau) \left(T\,e^{-i\int^\tau_{-\infty } d\tau'H_I(\tau')} \right)\rangle}
\begin{figure}[t!]
	\hfill
\begin{minipage}[t]{0.49\textwidth}
	\begin{flushleft}
		\epsfig{file=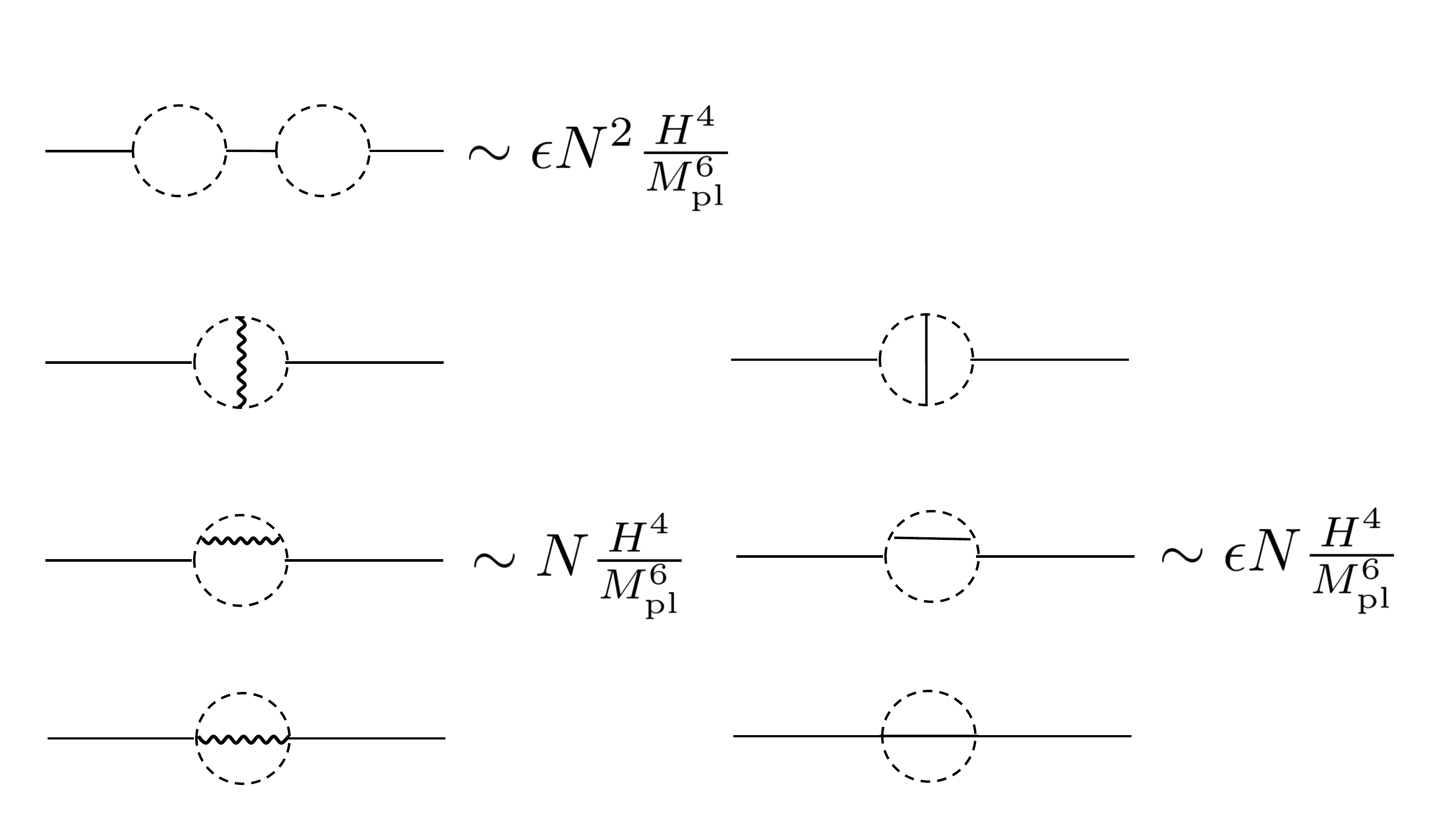, height=1.8in, width=3.3in}
		\caption{Two loop corrections to $\langle\zeta\zeta\rangle$. Wavy lines denote the graviton propagator. The double sunset graphs dominate when $N \gg 1/\epsilon$.
			\label{f:2loops}}
	\end{flushleft}
\end{minipage}
\hfill
	\begin{minipage}[t]{0.49\textwidth}
		\begin{flushleft}
			\epsfig{file=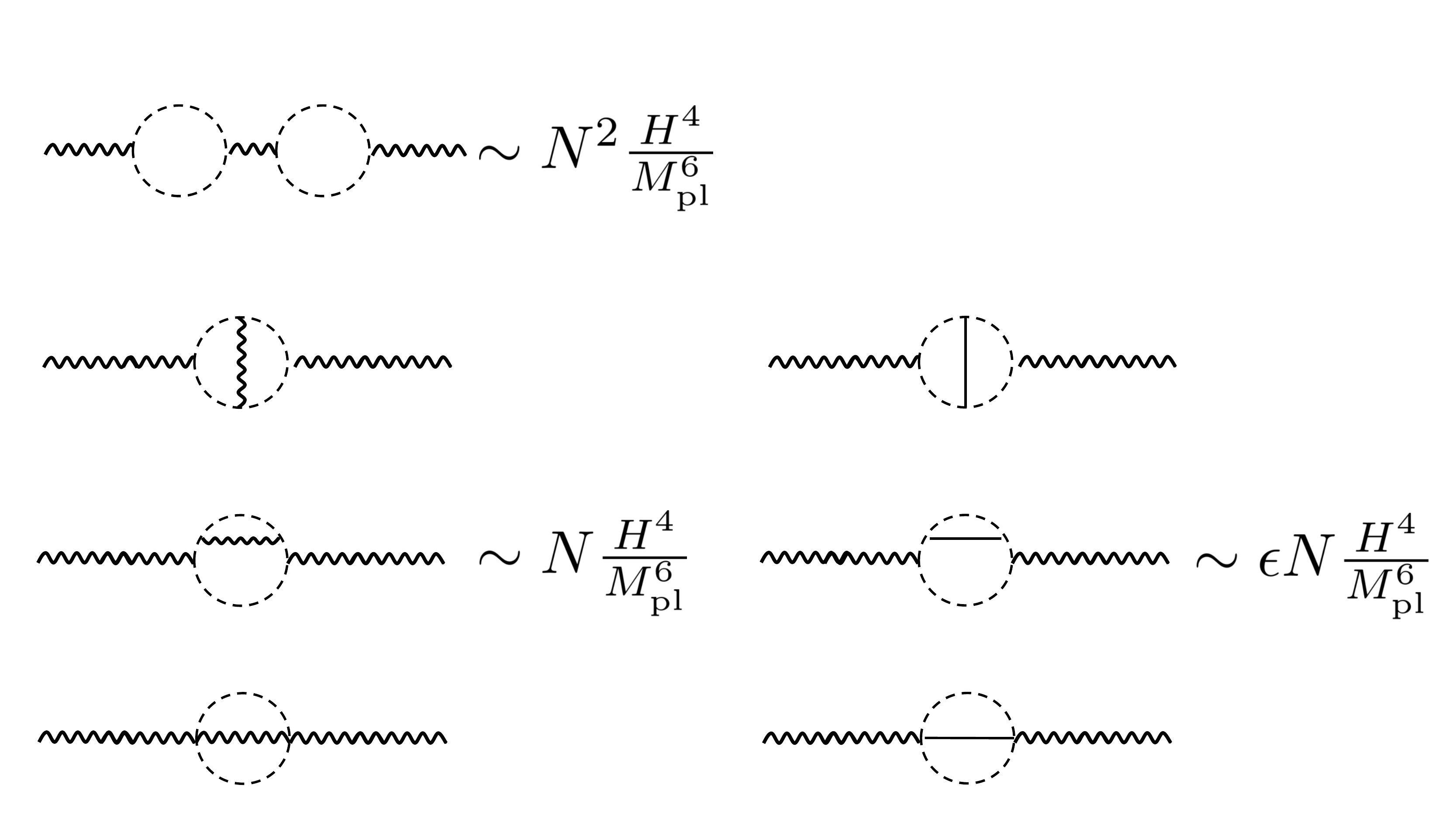, height=1.8in, width=3.3in}
			\caption{Two loop corrections to $\langle\gamma\gamma\rangle$, where here we only require $N \gg 1$ for the double sunset graphs to dominate.
				\label{f:2loopsg}}
		\end{flushleft}
	\end{minipage}
	\hfill
\end{figure}
Nominally the second order correction to (\ref{inin1}) is equivalent to the following expression \cite{weinberg}
\eq{sk1}{\langle \gamma^s_{ij, \K}(\tau)\gamma^{s'}_{ij,\Kp}(\tau) \rangle_{(2)} = -\int^\tau_{-\infty}d\tau_2\int^{\tau_2}_{-\infty}d\tau_1\langle[H_I(\tau_1),[H_I(\tau_2),\gamma^{0,s}_{ij, \K}(\tau)\gamma^{0,s'}_{ij,\Kp}(\tau)]] \rangle}
However, we have to be careful, since we shall be deforming the contour to imaginary time in the past in order to pick out the correct interacting vacuum. Therefore, we really need to be calculating
\eq{inin2}{\langle \gamma^s_{ij, \K}(\tau)\gamma^{s'}_{ij,\Kp}(\tau) \rangle = \langle \left(T\,e^{-i\int^\tau_{-\infty(1 + i\varepsilon) } d\tau'H_I(\tau')} \right)^\dag \gamma^{0,s}_{ij, \K}(\tau)\gamma^{0,s'}_{ij,\Kp}(\tau) \left(T\,e^{-i\int^\tau_{-\infty(1 + i\varepsilon')} d\tau'H_I(\tau')} \right)\rangle}
with $\varepsilon, \varepsilon'$ independent. This means that the symmetry in the domains of integration that allow one to express a time ordered product of integrals in terms of an integral over a simplex is broken whenever we have one operator from the time ordered product and another from the anti-time ordered product -- this is a requisite for the expression (\ref{inin}) to equal (\ref{sk1}), as first pointed out in \cite{Adshead}. Not accounting for this will result in missing contributions to the loop integral in addition to spurious divergences. Mindful of the latter, we go through the details of the calculation in appendix \ref{a:tensor}, considering additional subtleties arising from dimensional regularization on a quasi dS background. The intermediate result is the one loop correction
\eq{inter}{\calP_\gamma(k) = \frac{2H_*^2}{\pi^2\mpl^2}\left[1 + \frac{N}{16\pi^2} \frac{H_*^2}{\mpl^2}\frac{3}{5}{\rm log}\left(H_k/H_*\right)\right], }
where for the moment, we suppress slow roll corrections to the external mode functions that generate the usual tilt of the tensor power spectrum. We note that the $\log(k/\mu)$ dependence previously calculated in the literature (e.g. \cite{Adshead, weinberg}) is merely the first of multiple logarithmic corrections to the tree level result. An additive correction of the form $\log(-H_*\tau_k)$ also arises from corrections to the mode functions proportional to $\delta$ in $3 + \delta$ spatial dimensions \cite{SZ}, which goes over into a $\log(k/\mu)+ (1 + \epsilon)\log(-H_*\tau_k) = \log(H_k/\mu)$ dependence once one accounts for additional slow roll corrections (\ref{hk})\footnote{See the discussion in section 3.2 of \cite{SZ}, which we extend to quasi dS backgrounds in appendix \ref{a:tensor}.}. Upon fixing renormalized quantities at some pivot scale $H_*$, the dependence in (\ref{inter}) results. 

However, once one realizes that $H_k$ itself runs as inflation progresses, one finds that (cf. (\ref{htok}) and (\ref{sr2cf})) --
\eq{}{\log \frac{H_k}{H_*} = - \epsilon_* \log \frac{k}{k_*} + \mathcal O(\epsilon_*^2).}
Hence, the intermediate result for the one loop correction becomes:
\eq{interpt}{\calP_\gamma(k) = \frac{2H_*^2}{\pi^2\mpl^2}\left[1 - \epsilon_*\frac{N}{16\pi^2} \frac{H_*^2}{\mpl^2}\frac{3}{5}{\rm log}\left(k/k_*\right)\right]. }
That is, the net result of incorporating terms previously neglected in implementing dimensional regularization, whereby $\log(k/k_*)$ would have appeared in the intermediate expression\footnote{We note that the equivalent expression in \cite{Adshead} with a $\log k/\mu$ correction has a different numerical coefficient and opposite sign to that which would have appeared in (\ref{inter}), this error  has been acknowledged to us \cite{Peter}.} (\ref{inter}) instead of $\log(H_k/H_*)$, is to still induce a $\log k$ running, but of the opposite sign and with extra $\epsilon_*$ suppression. 
Additional corrections to the tensor two point function from the background dynamics suppressed in (\ref{interpt}) results in the final expression
\bea
\label{resumans4}
\calP_{\gamma}=\Delta_\gamma\left(\frac{k}{k_*}\right)^{-2\epsilon_* + \calO (\epsilon^2)} \left[ 1- \epsilon_* \frac{N}{16\pi^2} \frac{ H_*^2}{\mpl^2 }\frac{3}{5} \log \left(k/k_*\right) + \calO(\epsilon^2)\right] \, .
\eea
For completeness, we note that as illustrated for two loops in Fig. \ref{f:2loopsg}, in the limit $N \gg 1$, diagrams consisting of $n$ independent insertions of hidden loops dominate at the $n^{th}$ loop order and can in principle be resummed, allowing us to consistently infer the running if it is of the same order as that induced from the background alone. However, when doing this, one must be sure to have taken into account dependence on the slow roll parameters to all orders. Formally:
\bea
\label{resumans5}
\calP_{\gamma}=\frac{\Delta_\gamma\left(\frac{k}{k_*}\right)^{n_t(\epsilon_*,\dot\epsilon_*,...)}}{ 1+ \sim\hspace{-4pt}\bigcirc\hspace{-4pt}\sim} 
\eea
where $n_t(\epsilon_*,\dot\epsilon_*,...)$ denotes the spectral tilt to all orders in the Hubble hierarchy (cf. (\ref{nt}) to second order), and where
\eq{}{\sim\hspace{-4pt}\bigcirc\hspace{-4pt}\sim ~~=~~ \epsilon_{*}\frac{N}{16\pi^2} \frac{H_*^2}{\mpl^2 }\frac{3}{5} \log \frac{k}{k_{*}} + ...,  }
where the ellipses denote corrections to the running from non-trivial momentum dependent corrections to the mode functions, suppressed by extra factors of slow roll parameters that one can in principle calculate to the desired order (cf. appendix \ref{a:vert}), although in practice we shall only be interested in the leading order correction to the above.

In the limit $N \gg 1/\epsilon_* $ one can perform a similar resummation for the corrections to the power spectrum of the curvature perturbation, so that again formally
\bea
\label{resumans6}
\calP_{\zeta}=\frac{\Delta_\zeta\left(\frac{k}{k_*}\right)^{-1 + n_s(\epsilon_*,\dot\epsilon_*,...)}}{ 1 + -\hspace{-4pt}\bigcirc\hspace{-4pt}-  }
\eea
where
\eq{1lz}{-\hspace{-4pt}\bigcirc\hspace{-4pt}-~~=~~ -c\, \epsilon_*^2 \frac{N}{16\pi^2} \frac{H_*^2}{\mpl^2 } \log \frac{k}{k_{*}} + ...}
and where $n_s(\epsilon_*,\dot\epsilon_*,...)$ is the spectral tilt for the curvature perturbation. The ellipses again denote calculable corrections to the running from slow roll corrections the mode functions, and where the precise numerical coefficient $c$ (calculated to be $c = 4/15$ in \cite{weinberg}\footnote{\label{wfoot}We note that there is a spurious normalization factor of $1/(2\pi)^3$ in eq 72 relative to eq 71 of \cite{weinberg} given the conventions therein. Correcting for this results in a loop suppression factor of $\sim 1/(16\pi^2)$, consistent with our findings in appendix \ref{a:tensor}.}) in the above is unimportant to us\footnote{The $\epsilon_*$ suppression of the $\zeta$ vertices can never be compensated by large $N$ consistent with the strong coupling bound and will always result in loop corrections that are subleading to slow roll corrections from the background.} other than it positive, so that (\ref{1lz}) has an overall negative sign.

\section{Observational limits on the number of scalar fields $N$ \label{s3}}
Recalling the Hubble hierarchy of slow roll parameters
\eq{}{\epsilon_* = \epsilon_1 := - \frac{\dot H_*}{H_*^2},~~~ \epsilon_{i+1} := \frac{\dot \epsilon_i}{H_*\epsilon_i},}
one finds by taking appropriate logarithmic derivatives of the loop corrections to the scalar and tensor power spectra, additional corrections to the tilt and the running (to second order in the Hubble hierarchy \cite{Gong,Schwarz:2001vv,Casadio:2005xv})
\begin{eqnarray}
\label{ns}
n_s -1 &=& -2\epsilon_* - \epsilon_2 - 2\epsilon_*^2 - (2 C + 3)\epsilon_*\epsilon_2 - C\epsilon_2\epsilon_3 + c\,\epsilon_*^2\lambda\\
\label{as}
\frac{d n_s}{d \log k} &=& -2\epsilon_*\epsilon_2  - \epsilon_2\epsilon_3\\
\label{nt}
n_t &=& -2\epsilon_* + \epsilon_*\lambda - 2\epsilon_*^2 - 2(C + 1)\epsilon_*\epsilon_2\\
\label{at}
\frac{d n_t}{d \log k} &=& -2\epsilon_*\epsilon_2 + \frac{\lambda}{2}\epsilon_*\epsilon_2 - \frac{\epsilon_*^2}{2}\lambda^2
\end{eqnarray}
where $C = {\rm log}\,2 -2 + \gamma_E$, $\gamma_E$ being the Euler-Mascheroni constant and where we define
\eq{}{\lambda := \frac{3}{5}\frac{N}{16\pi^2}\frac{H_*^2}{\mpl^2} = \frac{3}{160}Nr_* \Delta_\calR.}
As reviewed in appendix \ref{a:sc}, we observe that $\lambda$ is necessarily bounded from above by 
\eq{ub}{\lambda \lesssim \frac{3}{5}} 
since $N H^2/(16 \pi^2\mpl^2) \lesssim 1$ in order for us to have a valid semi-classical calculation. Therefore the tilt and the running of the spectral index of the curvature perturbation is too feeble to compete with leading order corrections from the background (e.g. the correction to (\ref{as}) from loops of hidden fields begins at order $\epsilon_*^4\lambda^2$).

On the other hand, the spectral tilt for the tensor modes can receive corrections that are more cleanly observable:
\eq{nt2}{n_t = -2\epsilon_* + \epsilon_*\lambda.} 
Using the fact that the relation 
\eq{}{r_* = \frac{\calP_\gamma(k_*)}{\calP_\zeta(k_*)} = 16\epsilon_*}
remains unchanged, we see that the consistency relation is modified since we now have
\eq{stcr}{n_t = -\frac{r_*}{8}\left(1 - \frac{\lambda}{2}\right)}
which we rewrite as
\eq{}{ \frac{3.5\times 10^{11}}{r_*^2}\left(n_t + \frac{r_*}{8}\right) \approx N,}
where we have evaluated the numerical prefactor using the observed value $\Delta_\zeta \approx 2.44 \times 10^{-9}$. Therefore, if $(n_t + r_*/8)$ can be bounded from above by some positive number, i.e. if we can ever conclude to some threshold of confidence that 
\eq{}{n_t +  \frac{r_*}{8} \lesssim \xi} 
for some positive $\xi$, then one can bound
\eq{final}{N\lesssim \frac{\xi}{r_*^2}3.5\times10^{11}}
If we presume the most optimistic case that $r_* \sim 0.06$ then the best we can hope to bound $N$ is by
\eq{}{N \lesssim 10^{14}\times \xi}
Note that the strong coupling bound requires that
\eq{sc}{N \lesssim \frac{16\pi^2\mpl^2}{H^2}= \frac{32\Delta_\zeta^{-1}}{r_*} \approx \frac{1.3\times10^{10}}{r_*} }
so that in order to infer a stronger bound from (\ref{final}) than from consistency imposed by being below the strong coupling scale (\ref{sc}), we need to bound $\xi$ by an order of magnitude more accurately than the measured $r_*$. For CMB observations, this is on the threshold conceivable within cosmic variance limits -- at $r_* \sim \calO(10^{-1})$, the best one can hope to bound $\xi$ is approximately $\calO(10^{-2})$ \cite{Song:2003ca}, which is not much more constraining that the naive strong coupling bound. Whether future space based gravitational wave interferometry or ultimate 21 cm observations can improve upon this sensitivity is a possibility we contemplate in the following section. 

\section{Discussion\label{s:disc}}

For the purposes of the following, we frame the discussion in terms of an observational challenge for bounding $\xi$ in the context of (\ref{final}) as a null test. We will abuse our privileges as theorists to contemplate the possibility that one could bound $\xi$ past CMB cosmic variance limits to the level of $10^{-3}$ or beyond. That this may be plausible with a combination of future space based interferometry \cite{Danzmann:2003tv} and ground based arrays \cite{Janssen:2014dka} can be appreciated from the fact that the tilt for the tensor spectrum is no longer as negative as $\lambda$ approaches its upper bound (\ref{ub}), so that at comoving scales $k \sim 10^{14}k_* \sim 10^{11}$ Mpc$^{-1}$  (corresponding to peak interferometer sensitivities in the mHz range) the power will be enhanced by  about 20\% relative to the standard case. One might conceive improved prospects for constraining deviations from the consistency relation from combining observations sensitive enough to detect the stochastic primordial background \cite{Bartolo:2016ami} at widely separated scales, with CMB observations and space-based interferometry sensitive to modes 14 orders of magnitude apart\footnote{One might be concerned that higher order corrections might need to be incorporated in order to extrapolate the running over such a large range of scales. However, such corrections only become important when considering scales such that $\log k/k_* \sim \frac{1}{\lambda\epsilon_*} \gtrsim \frac{1}{\epsilon_*} = \frac{16}{r_*} \gtrsim 160$, which is safely beyond anything accessible to observations.}, with SKA like surveys interpolating between them with sensitivity at the nHz frequencies ($k \sim 10^{8}k_* \sim 10^{5}$ Mpc$^{-1}$). Ultimate 21 cm observations also offer the possibility to measure primordial gravitational wave background through large scale structure fossils, allowing for an in principle sensitivity to $r$ down to the $ \sim 10^{-6}$ level \cite{Masui:2010cz}, however, the question of whether foregrounds can be understood to the required level is far from settled at the present moment. For now, we merely state the obvious corollary that follows from (\ref{final}) that (for $r_* \sim 0.06$)
\eq{pmtest}{N  \lesssim \xi\cdot  10^{14} \sim 10^{9}-10^{12}}
for $\xi$ ranging from $10^{-5} \lesssim \xi \lesssim 10^{-2}$ where the latter corresponds to CMB cosmic variance bounds, and the former corresponds to us rather speculatively entertaining bounds that could be obtained by other means -- combinations of next generation space and ground based gravitational wave observations or ultimate 21cm observations. 

\subsection{Implications for BSM/ string models}
The idea of invoking a large number of hidden sectors to address the electroweak hierarchy problem was considered in \cite{Dvali:2007hz, Dvali:2007wp, Dvali:2009ne} -- the observation being that a large number of species can be used to make the scale of quantum gravity $\Lambda_{\rm QG}$ parametrically lower than the Planck mass (cf. appendix C)\footnote{\label{scg bound}This bound is often stated as $\Lambda_{\rm QG} \sim \frac{\mpl}{\sqrt N}$ where $\mpl$ is the reduced Planck mass and an order unity pre-factor is understood. As reviewed in the appendix, repeating the various arguments presented in \cite{Dvali:2007hz, Dvali:2007wp} suggests that the bound is \textit{at least} (\ref{lqg}).}
\eq{lqg}{\Lambda_{\rm QG} \sim \frac{4\pi\mpl}{\sqrt N}.}
Far from being an ad hoc construction, \cite{Dvali:2007wp, Dvali:2009ne} argue that such a large number of hidden sector arise naturally as Kaluza-Klein copies of the standard model in scenarios with extra dimensions, although their origin needn't be extra dimensional in general (see \cite{DSV} for an interesting speculation that these extra species could constitute dark matter). As discussed above, any observation of primordial tensors in the context of single field inflation immediately implies that in order for inflation to have occurred below the strong coupling scale, one must necessarily live in a universe with less than 
$N \lesssim \frac{16\pi^2\mpl^2}{H^2} \approx \frac{10^{10}}{r_*}$ hidden fields (\ref{sc}), with tests of deviations from the tensor to scalar consistency relation to better than the percent level allowing us more constraining power than the strong coupling bound. 

More recently, the authors of \cite{Arkani-Hamed:2016rle} proposed an alternative solution to the hierarchy problem that necessarily invokes inflation, initially dubbed `$N$-naturalness'. The idea is that we live in a universe with $N$ copies of the standard model each hidden from the other, with all coupled to a reheating field (the reheaton), not necessarily the inflaton. The mass of the Higgs fields in any of the $N$ copies of the standard model is drawn from a uniform distribution that interpolates between $-\Lambda^2 \leq m^2_H \leq \Lambda^2$. Given that reheating will preferentially produce particles in the lightest sector (with masses set by the Higgs expectation value), one dynamically explains why the universe that emerges from inflation will have a naturally small Higgs mass. A significant parameter space of interest lies within the range $N \sim 10^4 - 10^{16}$, for which tests of the tensor to scalar consistency relation at the per-mille level or better (\ref{pmtest}) could significantly constrain. 

\subsection{Generalization to higher spin}

In general, hidden sectors in string or BSM constructions possess a spectrum that is not restricted to scalar fields. An obvious question therefore is how our results generalize when including higher spins. Leaving aside the precise nature of the running of the two point function (the primary concern of this investigation), one can immediately infer the relative importance of the contributions from particles of different spins by consulting the one loop effective action obtained from integrating them out over a fixed background \cite{HK}. For a particle of a given spin, the effective action is given by (\ref{hkea}) (see appendix \ref{a:sc} for a discussion of the interpretation of the quantity below)
\eq{effact}{\calL_{\rm eff} = \frac{\mpl^2}{2}R + \frac{1}{2880\pi^2}\left[a_s R_{\mu\nu}R^{\mu\nu} + b_s R^2\right] + ...}
where the coefficients $a_s, b_s$ depend on the spin of the particle integrated out, and where we have used the Gauss-Bonnet relations in 4D to eliminate redundant operators. The relative contributions of the terms from which we have extracted our tree level result and our loop contributions can be determined from the ratio of the two contributions in (\ref{effact}). From the table in Appendix~\ref{a:sc} and eq.~(\ref{hkea}), we can read off the coefficients $a_s$ and $b_s$ for different spins to obtain for maximally symmetric backgrounds (where $R_{\mu\nu} = g_{\mu\nu}R/4$)
\begin{eqnarray}
\label{spinw}
\frac{\calL_{\rm 1-loop}}{\calL_{\rm tree}} &=& \frac{9/4}{1440\pi^2}\frac{R}{\mpl^2};~~ {\rm spin~ 0},\\ \nn 
&=& \frac{3/2}{1440\pi^2}\frac{R}{\mpl^2};~~ {\rm spin~ 1/2},\\ \nn
&=& \frac{-3}{1440\pi^2}\frac{R}{\mpl^2};~~ {\rm spin~ 1}.
\end{eqnarray}
From this, we can conclude that in a universe with a spectrum of particles consisting of $N_\phi$ scalars, $N_\psi$ Dirac fermions and $N_V$ $U(1)$ gauge fields, the actual quantity one is bounding with (\ref{final}) is the relevant spin weighted sum indicated in (\ref{spinw}). We leave the explicit computation of this index and the running induced by higher spin fields for a future study \cite{DDP}.

We note in passing that on a dS background, $R = 12H^2$, so that for $N$ scalar fields we have 
\eq{}{2\times \frac{\calL_{\rm 1-loop}}{\calL_{\rm tree}} = \frac{2N\cdot 9/4}{1440\pi^2}\frac{R}{\mpl^2} = \frac{N}{16\pi^2}\frac{3}{5}\frac{H^2}{\mpl^2}}
where the factor two is to count the two independent polarizations that contribute to the tensor power spectrum. This is exactly the relative ratio of the loop contribution to the tree level result calculated in  (\ref{resumans4}), providing a non-trivial check on our results. 

\subsection{Possible implications for eternal inflation}

Although not the primary focus of this investigation, having to come to terms with the precise nature of the slow roll corrections to the loop integrals (and correctly implementing dimensional regularization on a quasi dS spacetime) has potential implications for eternal inflation\footnote{Recently, the authors of \cite{Agrawal:2018own} have proposed a conjecture motivated from string `swampland' considerations \cite{Obied:2018sgi} that suggest obstacles for accomplishing inflation at all within string theory. Insofar as our study takes a viable inflating background for granted, the presence of additional hidden fields with no potential terms is no more problematic than assuming an inflationary background in the first place. }. Recalling the discussion of \cite{SZ}, who discovered by using a \textit{mass dependent} regularization scheme (a hard cutoff in physical momentum) that logarithmic corrections to the the two point function of the curvature perturbation of the form $\log H_*/\mu$ resulted. This was in contradiction with the $\log k/\mu$ form of the loop correction derived elsewhere in the literature when applying dimensional regularization. Senatore and Zaldarriaga reasoned that the former could not be the final answer -- taking the result \cite{weinberg} at face value,
\eq{}{\calP_\zeta = \frac{H^2_*}{8\pi^2\mpl^2\epsilon_*}\left[1  - \epsilon_* \frac{N}{16\pi^2} \frac{4}{15}\frac{H_*^2}{\mpl^2 } \log \frac{k}{\mu}\right]}
one finds upon Fourier transforming back to position space, that the variance of the inflaton fluctuation $\delta \phi = -\frac{\dot{\phi_0}}{H_*}\zeta$ is given by
\eq{weinc}{\langle \delta\phi^2 \rangle_{{\rm 1-loop}} = - \frac{\epsilon_* N}{15\pi} \frac{H_*^2}{\mpl^2 }\int^{\Lambda a(t)} d^3k \frac{H^2_*}{k^3}\log k \sim   - \frac{\epsilon_* N}{15\pi} \frac{H_*^2}{\mpl^2 } H_*^2 (\log a)^2 \sim - \frac{\epsilon_* N}{15\pi} \frac{H_*^2}{\mpl^2 } H_*^4 t^2  }
implying that the fluctuations of the inflaton field decay monotonically over time, implying that \textit{no model of inflation is eternal} were the form of the correction (\ref{weinc}) to be trusted\footnote{A more thorough treatment is provided in \cite{eternal} where it is shown that the reheating volume diverges above the critical inflaton velocity $\dot\phi^2_{\rm cl}/H^4 > 3/(2\pi^2)$. Here, $\phi_{\rm cl}$ is to be understood as the field around which one implements background field quantization -- i.e. the field that minimizes the \textit{effective action}. For the purposes of the present discussion, we content ourselves with observing the growth or decay of the variance, which after repeating the steps of \cite{eternal} can be shown to result in crossing this critical velocity or not.}, which clearly cannot be the case. 

The resolution pointed out by \cite{SZ} was that the $\log k/\mu$ corrections found previously were merely the first of several logarithmic contributions that had to be supplemented with corrections to the dimensionally deformed mode functions and integration measures that went as $\log(-H_*\tau_k)$, where $\tau_k$ is the time of crossing of the comoving $k$-mode. Adding up all such corrections resulted in a dependence of the form
\eq{}{\log k/\mu + \log (-H_*\tau_k) = \log H_*/\mu,} 
in agreement with the results obtained with a hard cutoff, which suggest that the correlation functions do not run at this order. However as we have shown, this is not the final story either, since the only possibility for  which correlation functions of an interacting theory remain independent of scale is if we are at a fixed point of the theory, where a scale symmetry is realized. This is indeed the case in the strict dS limit, where $H$ is constant and one has an exact dilatation invariance. Therefore, moving away from the strict dS limit must reintroduce a running calculated in appendix B (to next to leading order in slow roll)  with the result (\ref{hk}):
\eq{}{\log k/\mu + (1 + \epsilon)\log (-H_*\tau_k) = \log H_k/\mu,}
Upon fixing renormalization conditions at a particular pivot scale $\mu = H_*$, this implies that correlation functions will run as (\ref{htok})
\eq{}{\log \frac{H_k}{H_*} = -\epsilon \log \frac{k}{k_*}}
Therefore, repeating the calculation for the corrections to the curvature two point function, one finds that a $\log k$ running is reintroduced, but of the opposite sign and with additional slow roll suppression. Retracing the argument leading to (\ref{weinc}) seems to imply that our results imply that all models of inflation in the presence of hidden fields are eternal. However this is too naive, as we have to factor in corrections from the background as well. If the sign of the net log correction $\log H_k/H_*$ generated from background corrections and cubic self interactions of the curvature perturbation alone were always positive, one would then be able to conclude that \textit{indeed, all models of inflation were eternal}, as argued to be the case in \cite{Ijjas:2015hcc}. However, it is very likely that loop corrections do not always compete with classical logarithmic corrections from the background evolution, and the precise conclusion one arrives at depends on the given background model. This is an important issue which deserves a thorough separate investigation.

\section*{Acknowledgements}
We thank Peter Adshead, Cliff Burgess, Martin Sloth, Filippo Vernizzi and Matias Zaldarriaga for valuable and informative discussions over the course of this investigation. We acknowledge support from the Swiss National Science Foundation. SP is supported by funds from Danmarks Grundforskningsfond under grant no. 1041811001. A. d. R. is supported by the Spanish  Ph.D fellowship FPU13/04948 and research stay Grant EST15/00296, and thanks the Theoretical Cosmology Group of the University of Geneva for hospitality during the initial stages of this collaboration.

\begin{appendix}
	\section{On the $\epsilon$ dependence of the $\zeta$ vertices}\label{a:vert}
We consider the action for the zero mode of the (canonically normalized) inflaton plus N hidden scalars.
\begin{eqnarray}
\label{act}S &=& \frac{\mplf^2}{2}\int d^4x\sqrt{-g}R[g] - \frac{1}{2}\int d^4x\sqrt{-g}\left[\partial_\mu\phi\partial^\mu\phi + 2V(\phi) \right]\\\nonumber &-&  \sum_{n=1}^{n_{\rm max}} \frac{1}{2}\int d^4x\sqrt{-g}\,\partial_\mu\chi_n\partial^\mu\chi_n
\end{eqnarray}
By assumption, the $\chi$ fields have no classically evolving background, and so appear in the action to leading order as quadratic in perturbations. We ADM decompose the metric
\eq{adm}{ds^2 = -N^2dt^2 + h_{ij}(dx^i + N^idt)(dx^j + N^jdt),}
and work in comoving gauge
\bea
\label{cgf}
\phi(t,x) &=& \phi_0(t),\\
h_{ij}(t,x) &=& a^2(t)e^{2\calR(t,x)}\delta_{ij}.
\eea
This gauge is defined by the foliation where the inflaton is the clock (no other field has a background). Writing
\bea
\label{exp}
N &=& 1 + \alpha_1\\ \nn
N^i &=& \partial_i\theta + N^i_T~,~ \mbox{ with }~\partial_i N^i_T\equiv 0
\eea
where $\alpha_1, \theta$ and $N^i_T$ all first order quantities, we find the solutions (we only need to calculate to first order for the constraints to obtain the action to cubic order \cite{Maldacena})
\eq{asolc}{\alpha_1 = \frac{\dot \calR}{H}}
\eq{tsolc}{\partial^2\theta = -\frac{\partial^2\calR}{a^2 H} + \epsilon \dot\calR }
	where $\partial^2 = \partial_i\partial_i$ contains no factors of the scale factor, and where $\epsilon$ is defined as:
	\eq{eps}{\epsilon := \frac{\dot\phi_0^2}{2H^2\mplf^2}}
	The relevant quadratic and cubic terms are (summation over $n$ implicit)
	\eq{s2r}{S_{2,\calR} = \mplf^2\int d^4x\, a^3\,\epsilon\left[\dot\calR^2 - \frac{1}{a^2}(\partial\calR)^2 \right]}
	\eq{s2h}{S_{2,\chi} = \frac{1}{2}\int d^4x\, a^3\left[\dot\chi_n{\dot\chi_n} - \frac{1}{a^2}\partial_i\chi_n\partial_i\chi_n\right]}
	
	\begin{eqnarray}
	\label{s3h} S_{3,\calR \chi} = \frac{1}{2}\int d^4x\Biggl\{&&\hspace{-15pt} a^3\dot\chi_n{\dot\chi_n}\left(3\calR - \frac{\dot\calR}{H}\right) - 2 a^3\dot\chi_n\partial_i\theta\partial_i\chi_n- a^3\left(\calR + \frac{\dot\calR}{H}\right)\frac{1}{a^2}\partial_i\chi_n\partial_i\chi_n\Biggr\}
	\end{eqnarray}
	We do not write the cubic action for $\calR$ since it all we shall need from it is the fact that it is suppressed by $\epsilon^2$ to leading order after enough integrations by parts \cite{Maldacena}. A similar thing happens for (\ref{s3h}) -- although it may appear that the cubic interactions between $\calR$ and the $\chi^a$ might be of order $\epsilon^0$, these interactions in fact of order $\epsilon$. This is readily seen by realizing that this contribution to the action is nothing other than the variation of the quadratic action for the hidden fields to first order in metric perturbations. That is, if $\calL_\chi = -\frac{1}{2}\partial_\mu\chi_n\partial^\mu\chi_n$, then the cubic interaction action for the hidden fields is given merely by the first order variation 
	\eq{cubact}{S_{3, \calR \chi} = \delta_{g_{\mu\nu}}\int \sqrt{-g}\calL_\chi = \frac{1}{2}\int\sqrt{-g}\,T^{\mu\nu}_{\chi}\delta_1 g_{\mu\nu},}
	where $T_\chi^{\mu\nu}$ is given by
	\eq{tdef}{T^{\mu\nu}_\chi = \left[ -\frac{g^{\mu\nu}}{2}\partial_\lambda\chi_n\partial^\lambda\chi_n + \partial^\mu\chi_n\partial^\nu\chi_n\right]}
	From (\ref{adm}), (\ref{asolc}) and (\ref{tsolc}) we see that the first order metric variations can be read off  as 
	\eq{1o}{\delta_1g_{\mu\nu} = \begin{pmatrix} -2\frac{ \dot\calR}{H}&&a^2\partial_i\theta\\a^2\partial_i\theta&& a^2\delta_{ij}2\calR \end{pmatrix}}
	One can explicitly verify that the trace of the product of the above with (\ref{tdef}) reproduces (\ref{s3h}). We observe that one can write (\ref{1o}) as 
	%
	\eq{1or}{\delta_1 g_{\mu\nu} = \nabla_\mu\beta_\nu + \nabla_\nu\beta_\mu + \Delta_{\mu\nu},}
	where 
	\eq{eodef}{\beta_0 = -\frac{\calR}{H},~\beta_i \equiv 0} 
	and where 
	\eq{ddef}{\Delta_{\mu\nu} := \epsilon \begin{pmatrix} 2\calR&&a^2\partial_i\partial^{-2}\dot\calR\\a^2\partial_i\partial^{-2}\dot\calR&& 0 \end{pmatrix}}
	Clearly only the second term in (\ref{1or}) gives a non-vanishing contribution. Therefore the relevant cubic interactions are given by
	\eq{cubint}{S_{3, \calR \chi} = \int d^4x\,a^3  \epsilon\left[\frac{\calR}{2}\left(\dot\chi_n{\dot\chi^n} + \frac{1}{a^2}\partial_i\chi_n\partial_i\chi_n\right) - \dot\chi_n\partial_i\chi_n\partial_i\partial^{-2}\dot\calR\right] }
	where we take note of the advertised $\epsilon$ suppression of the cubic interaction vertices. The relevant cubic interaction term for the tensor perturbations can be read off straightforwardly as
\eq{}{S_{3, \gamma \chi} = \frac{1}{2}\int d^4x\,a\left[\gamma_{ij}\partial_i\chi_n\partial_j\chi_n \right].}
For completeness, we note that the $\epsilon$ suppression accompanying factors of $\zeta$ extends to mixed vertices as well. For instance, consider the quartic $\zeta\gamma\chi\chi$ interaction obtained from expanding (\ref{act}) and solving for the lapse and shift constraints
\eq{}{S_{4, \gamma \chi \zeta} = \frac{1}{2}\int d^4x\,a\left(\frac{\dot\zeta}{H} + \zeta\right)\left[\gamma_{ij}\partial_i\chi_n\partial_j\chi_n \right].}
A straightforward integration by parts of the first term in the round parenthesis brings the above to the form
\eq{quint}{S_{4, \gamma \chi \zeta} = -\frac{1}{2}\int d^4x\,a\,\epsilon\, \zeta\left[\gamma_{ij}\partial_i\chi_n\partial_j\chi_n \right] + ...  }
with the slow roll suppression now manifesting, and where the ellipses denote terms proportional to $\partial_t \left[\gamma_{ij}\partial_i\chi_n\partial_j\chi_n \right]$ which can be eliminated via a suitable field redefinition using the background equations of motion similar to the ones considered in \cite{Maldacena}.
	
\section{One loop correction to $\langle\gamma\gamma\rangle$}\label{a:tensor}
Mindful of the cautions articulated in \cite{Adshead}, we are interested in expanding (\ref{inin2}) to second order:
\eq{ainin}{\langle \gamma^s_{ij, \K}(\tau)\gamma^{s'}_{ij,\Kp}(\tau) \rangle = \langle \left(T\,e^{-i\int^\tau_{-\infty(1 + i\varepsilon) } d\tau'H_I(\tau')} \right)^\dag \gamma^{0,s}_{ij, \K}(\tau)\gamma^{0,s'}_{ij,\Kp}(\tau) \left(T\,e^{-i\int^\tau_{-\infty(1 + i\varepsilon') } d\tau'H_I(\tau')} \right)\rangle,}
where we have switched to conformal time, with the interaction Hamiltonian given by
\eq{b2}{H_I = -\sum_r\frac{1}{2}\int d^3x\,a^2\left[\gamma^{0,r}_{ij}\partial_i\chi_n\partial_j\chi_n \right],}
where a sum over $n$ is understood. Consistent with the normalizations (\ref{psdef}) and (\ref{pstdef}), the free fields admit the expansion
\eq{}{\chi^a(\X,\tau) = \int\frac{d^3k}{(2\pi)^3} e^{i\K\cdot\X}\left[b^a_{\K}\chi_{\K}(\tau) + b^{a\dagger}_{-\K}\chi^*_{\K}(\tau) \right] }
\eq{}{\gamma^{0,r}_{ij}(\X,\tau) = \int\frac{d^3k}{(2\pi)^3} e^{i\K\cdot\X}\left[\epsilon^r_{ij}(\K) a^r_{\K}\gamma_{\K}(\tau) + \epsilon^{r*}_{ij}(-\K)a^{r \dagger}_{-\K}\gamma^*_{\K}(\tau) \right] }
with creation and annihilation operators normalized as
\eq{opnorm}{\left[a^s_\K, a^{r\dagger}_{\Q}\right] = \left(2\pi\right)^3\delta^{sr}\delta^{3}\left(\K - \Q\right);~~\left[b^a_\K, b^{b\dagger}_{\Q}\right] = \left(2\pi\right)^3\delta^{ab}\delta^{3}\left(\K - \Q\right)}
and where the polarization tensors are normalized as
\eq{b6}{ \epsilon^r_{ij}(\K)\epsilon^{*s}_{ij}(\K) = 4\delta^{rs};~~{\rm with}~~ \epsilon^{r*}_{ij}(\K) = \epsilon^r_{ij}(-\K)}
with mode functions $\chi_{\K}$ and $\gamma_{\K}$ canonically normalized such that in the dS limit, they're given by
\eq{wfc}{\chi_{\K}(\tau) = \frac{iH}{\sqrt{2k^3}}\left(1 + i k\tau\right)e^{-i k\tau} }
\eq{wfg}{\gamma_{\K}(\tau) = \frac{iH}{\sqrt{2k^3}\mplf}\left(1 + i k\tau\right)e^{-i k\tau}.}
We note that there is one term with two interaction operators from the time ordered product in (\ref{ainin}) plus one term with two operators from the anti-time ordered product, plus one term with interaction operators from both the time ordered and the anti-time ordered products. The latter can be written as term II below, whereas the first two are complex conjugates of each other, and because of the fact that both contain the same lower limits, can be written as twice the integral over a triangle, which we group together to form term I:
\begin{eqnarray}
\nn
\langle \gamma^s_{ij, \K}(\tau)\gamma^{s'}_{ij,\Kp}(\tau) \rangle_{(2)} &=& -2\Re \int^\tau_{-\infty_+}d\tau_2\int^{\tau_2}_{-\infty_+}d\tau_1 \langle \gamma^{0,s}_{ij, \K}(\tau)\gamma^{0,s'}_{ij,\Kp}(\tau)H_I(\tau_2)H_I(\tau_1) \rangle~~~{\rm (I)}\\ &+& \int^\tau_{-\infty_+}d\tau_2\int^{\tau}_{-\infty_-}d\tau_1 \langle H_I(\tau_1) \gamma^{0,s}_{ij, \K}(\tau)\gamma^{0,s'}_{ij,\Kp}(\tau)H_I(\tau_2)\rangle~~~~~~~~~{\rm (II)}
\end{eqnarray}
where $\infty_\pm := \infty(1 \pm i\varepsilon)$. We note that (II) has different lower limits but identical upper limits, and will turn out to be an absolute value. We insert into the above the expansion (and dropping the zero superscripts to denote free field operators)
\eq{}{H_I(\tau_1) = -\frac{1}{2}\sum_r \int \frac{d^3k_1\,d^3p_1}{(2\pi)^6}a^2(\tau_1)\gamma^r_{lm,\K_1}(\tau_1)\chi^a_{\Pb_1}(\tau_1)\chi^a_{-\K_1-\Pb_1}(\tau_1)p_{1\,l}(k_{1\,m} + p_{1\,m}) }
and similarly for $H_I(\tau_2)$, where
\eq{}{\chi^a_{\K}(\tau) = b^a_{\K}\chi_{\K}(\tau) + b^{a\dagger}_{-\K}\chi^*_{\K}(\tau),}
with creation and annihilation operators normalized as per (\ref{opnorm}),  $\chi^a_\K$ the mode functions (which depend only on the magnitude of $\K$) and where
\eq{}{\gamma^r_{lm,\K}(\tau) =  \epsilon^r_{lm}(\K) a^r_{\K}\gamma_{\K}(\tau) + \epsilon^{r*}_{lm}(-\K)a^{r \dagger}_{-\K}\gamma^*_{\K}(\tau).}
In the basis where the graviton propagates along the z-direction, the polarization tensor corresponding to the normalization (\ref{b6}) is given by
\eq{epsnorm}{\epsilon^+_{lm}(\K) = \begin{pmatrix}1 && i && 0\\ i&& -1&& 0\\ 0&&0&&0 \end{pmatrix},}
where the index $r = \pm$. After Wick contracting, and utilizing the relations (\ref{b6}) and (\ref{epsnorm}) to do the contractions with the remaining momenta, we find (working around a dS background with $a = -(H\tau)^{-1}$) that term (II) is given by
\begin{eqnarray}
{\rm (II)} &=& 2N\frac{\delta^{s s'}}{H^4}\delta^3(\K + \Kp)\int^\tau_{-\infty_+}\frac{d\tau_2}{\tau_2^2}\int^{\tau}_{-\infty_-}\frac{d\tau_1}{\tau_1^2}\int d^3q\, q^4\,{\rm sin^4}\theta\\ \nn &\times& \Biggl\{ \chi_{\Q}(\tau_1)\chi^*_\Q(\tau_2)\chi_{\Q-\K}(\tau_1)\chi^*_{\Q+ \Kp}(\tau_2)\gamma_{\K}(\tau_1)\gamma^*_\K(\tau)\gamma_\Kp(\tau)\gamma^*_\Kp(\tau_2) \\ \nn &&+ \chi_{\Q}(\tau_1)\chi^*_\Q(\tau_2)\chi_{\Q-\Kp}(\tau_1)\chi^*_{\Q+ \K}(\tau_2)\gamma_{\Kp}(\tau_1)\gamma^*_\Kp(\tau)\gamma_\K(\tau)\gamma^*_\K(\tau_2)\Biggr\} \,.
\end{eqnarray}
This expression can be taken on shell as far as the wavefunctions are concerned, and using the explicit forms (\ref{wfc}) and (\ref{wfg}), we end up with the expression
\begin{eqnarray}
{\rm (II)} &=& \nn \frac{N}{4}\frac{H^4}{\mpl^4}\delta^{s s'}\delta^3(\K + \Kp)(1 + k^2\tau^2)\int\,d^3q\int\,d^3\bar q \,\delta^3(\bar\Q + \Q + \K) \frac{q\,{\rm sin^4}\theta}{k^6\bar q^3}\\&&\times \Bigg|\int^{\tau}_{-\infty_+}\frac{d\tau_1}{\tau_1^2} (1 + iq \tau_1)(1 + i\bar q \tau_1)(1 + i k \tau_1)e^{-i\tau_1(q + \bar q + k)} \Bigg|^2\label{II}
\end{eqnarray}
which results in, after taking the $\tau \to 0$ limit, the expression
\eq{IIans}{{\rm (II)} = \frac{N}{4}\frac{H^4}{\mpl^4}\delta^{s s'}\delta^3(\K + \Kp)\int\,d^3q\int\,d^3\bar q \,\delta^3(\bar\Q + \Q + \K) \frac{q\,{\rm sin^4}\theta}{k^6\bar q^3}\left[\frac{1}{\tau^2} + \Gamma^2 - 2\alpha + k^2\right]}
where
\eq{gamma}{\Gamma := \frac{(k^2 + q\bar q)(q + \bar q) + k(q^2 + \bar q ^2 + 4q\bar q ) }{(k + q + \bar q)^2}}
and where we shall also define for future convenience
\eq{alpha}{\alpha := \frac{kq\bar q}{(k + q + \bar q)}} 
The apparently divergent term in the $\tau \to 0$ limit in (\ref{IIans}) will cancel a corresponding term from (I), which we turn towards now. As before, inserting the expression for $H_I$ and performing the relevant Wick contractions and traces over polarization indices, one ends up with
\begin{eqnarray}
{\rm (I)} &=& \nn -\frac{N}{2}\frac{H^4}{\mpl^4}\delta^{s s'}\delta^3(\K + \Kp)\int\,d^3q\int\,d^3\bar q \,\delta^3(\bar\Q + \Q + \K) \frac{q\,{\rm sin^4}\theta}{k^6\bar q^3}\\&&\nn \times \Re\Biggl\{(1 - ik\tau)^2 e^{2ik\tau}\int^{\tau}_{-\infty_+}\frac{d\tau_2}{\tau_2^2} (1 - iq \tau_2)(1 - i\bar q \tau_2)(1 + i k \tau_2)e^{i\tau_2(q + \bar q - k)}\\&& ~~~~~~~~~~~~~~~~~~~~~~~~~~\int^{\tau_2}_{-\infty_+}\frac{d\tau_1}{\tau_1^2} (1 + iq \tau_1)(1 + i\bar q \tau_1)(1 + i k \tau_1)e^{-i\tau_1(q + \bar q + k)} \Biggl\}\label{I}
\end{eqnarray}
Performing the $\tau_1$ integral results in the intermediate expression
\begin{eqnarray}
{\rm (I)} &=&  -\frac{N}{2}\frac{H^4}{\mpl^4}\delta^{s s'}\delta^3(\K + \Kp)\int\,d^3q\int\,d^3\bar q \,\delta^3(\bar\Q + \Q + \K) \frac{q\,{\rm sin^4}\theta}{k^6\bar q^3}\\&&\nn \times \Re\Biggl\{(1 - i k\tau)^2e^{2 i k \tau}  \int^{\tau}_{-\infty_+}\frac{d\tau_2}{\tau_2^2} (1 - iq \tau_2)(1 - i\bar q \tau_2)(1 + i k \tau_2)e^{-2i\tau_2k}\left(-\frac{1}{\tau_2} + \alpha\tau_2 - i\Gamma\right) \Biggl\}
\end{eqnarray}
Performing the $\tau_2$  integral and taking the $\tau \to 0$ limit one ends up with
\begin{eqnarray}
{\rm (I)} &=& -\frac{N}{4}\frac{H^4}{\mpl^4}\delta^{s s'}\delta^3(\K + \Kp)\int\,d^3q\int\,d^3\bar q \,\delta^3(\bar\Q + \Q + \K) \frac{q\,{\rm sin^4}\theta}{k^6\bar q^3}\\ \nn&&\Biggl\{\frac{1}{\tau^2} - \frac{1}{2k^2}\left(2k^4 - 3k\alpha(q + \bar q) + 2q\bar q\alpha + 8k^3(q + \bar q - \Gamma) + 3kq\bar q\Gamma + 2k^2(q\bar q + \alpha - (q + \bar q)\Gamma) \right)   \Biggl\}
\end{eqnarray}
We note that when one evaluates this integral, there are nominal terms that go like ${\rm log~ k}$ and ${\rm log~ \tau}$, however these multiply a term whose coefficients cancel out once imposing the relations (\ref{gamma}) and (\ref{alpha}). Adding (I) and (II), the divergent term in $1/\tau^2$ cancels, leaving us with
\begin{eqnarray}
\label{IpII}
{\rm (I) + (II)} &=& \frac{N}{4}\frac{H^4}{\mpl^4}\frac{\delta^{s s'}}{k^6}\delta^3(\K + \Kp)\int\,d^3q\int\,d^3\bar q \,\delta^3(\bar\Q + \Q + \K) \frac{q\,{\rm sin^4}\theta}{\bar q^3}\\ \nn&&\Biggl\{ 2 k^2 - \alpha + \frac{q\bar q\alpha}{k^2} + q(\bar q - \Gamma) + 4k(q + \bar q - \Gamma) - \bar q\Gamma + \Gamma^2 - \frac{3(\bar q\alpha + q(\alpha - \bar q\Gamma))}{2k}   \Biggl\}
\end{eqnarray}
We now make use of the identity 
\eq{iden}{\int d^3q\,d^3\bar q\,\delta^3(\Q + \bar\Q + \K)f(q,\bar q,k) = \frac{2\pi}{k}\int_0^\infty dq\,q\int^{q + k}_{|q-k|} d\bar q\,\bar q\, f(q,\bar q,k) }
which can be seen by first doing the $\bar q$ integral with the delta function, leaving us with $2\pi\int d({\rm cos}\theta_q)\,q^2\,dq$, and using the relation 
\eq{cosrel}{\bar q^2 = q^2 + k^2 + 2qk\,{\rm cos}\,\theta_q}
to express $d\,{\rm cos}\,\theta_q = \bar q d\bar q/(qk)$ to arrive at the above. One can also use (\ref{cosrel}) to write
\eq{}{{\rm sin^4}\theta = \frac{\left(4q^2k^2 - (\bar q^2 - q^2 - k^2)^2 \right)^2}{16q^4k^4}}
Therefore in (\ref{iden}), the integrand is
\begin{eqnarray}
\label{fdef}
f(q,\bar q, k) &=& \frac{\left(4q^2k^2 - (\bar q^2 - q^2 - k^2)^2 \right)^2}{16q^4k^4} \frac{q}{\bar q^3} \times \nonumber \\
&& \hspace{-1.5cm}\Biggl\{ 2 k^2 - \alpha + \frac{q\bar q\alpha}{k^2} + q(\bar q - \Gamma) + 4k(q + \bar q - \Gamma) - \bar q\Gamma + \Gamma^2 - \frac{3(\bar q\alpha + q(\alpha - \bar q\Gamma))}{2k}    \Biggl\} 
\end{eqnarray}
At this stage, we need to evaluate the integral (\ref{iden}) via dimensional regularization. However there are various subtleties one must keep track in doing this correctly (elaborated upon in \cite{SZ}) which we address in what follows.

\subsection{Dimensional Regularization on a dS background}
In order to dimensionally regularize the loop integral (\ref{iden}), we realize on dimensional grounds that 
\eq{drint}{\int d^3q\,d^3\bar q\,\delta^3(\Q + \bar\Q + \K)f(q,\bar q,k) = k^{3+\delta}F(\delta)}
where $F(\delta)$ is a dimensionless constant which admits an expansion in powers of $\delta$
\eq{}{F(\delta) = \frac{F_0}{\delta} + F_1 + ...}
so that in the limit $\delta \to 0$, we find
\eq{drfint}{\int d^3q\,d^3\bar q\,\delta^3(\Q + \bar\Q + \K)f(q,\bar q,k) = k^{3}\left(F_0\, {\rm log}\,k + \Lambda\right)}
with $\Lambda$ a divergent constant to be subtracted with appropriate counter-terms. Hence, multiplying the integral by $k$ and taking the fifth derivative w.r.t. $k$ implies that the right hand side is given by $24 F_0/k$ and the left hand side is given by evaluating (\ref{iden}) using (\ref{fdef}) to result in $12\pi\cdot 24/(5k)$, so that 
\eq{}{F_0 = \frac{12\pi}{5}.}
From this we conclude the contribution
\eq{}{\frac{k^3}{2\pi^2}\langle \gamma^s_{ij, \K}(\tau)\gamma^{s'}_{ij,\Kp}(\tau) \rangle_{(2)} \supset \frac{N}{8\pi^2}\frac{H^4}{\mpl^4}\delta^{ss'}\delta^{3}(\K + \Kp)\left(\frac{12\pi}{5}\log(k/\mu) + \Lambda\right).}
However, as pointed out in \cite{SZ}, we are not done yet. As elaborate upon further in the next subsection, there are additional contributions coming from corrections to the dS mode functions themselves in $D = 3 + \delta$ dimensions:
\eq{drc}{\chi_\K(\tau) = -\frac{\sqrt{\pi}}{2}e^{i\pi\delta/4}\frac{H^{1 + \delta/2}}{\mu^{\delta/2}} \frac{(-k\tau)^{(3 + \delta)/2}}{k^{(3 + \delta)/2}}H^{(1)}_{(3 + \delta)/2}(-k\tau),}
\eq{drg}{\gamma_\K(\tau) = -\frac{\sqrt{\pi}}{2}e^{i\pi\delta/4}\frac{H^{1 + \delta/2}}{\mpl\,\mu^{\delta/2}} \frac{(-k\tau)^{(3 + \delta)/2}}{k^{(3 + \delta)/2}}H^{(1)}_{(3 + \delta)/2}(-k\tau).}
Expanding in $\delta$ results in (for example)
\eq{wfcdr}{\chi_{\K}(\tau) = \frac{iH}{\sqrt{2\mu^\delta k^3}}\left(1 + i k\tau\right)e^{-i k\tau}\left[1 + \frac{\delta}{2}\log(-H\tau) + 
	\frac{\delta}{2} u(-k\tau) + ...\right] }
where $u(-k \tau)$ is given by \cite{Chluba}
\eq{ukt}{u(-k\tau) = \left( \left[{\rm Ci}(-2k\tau) + i{\rm Si}(-2k\tau)\right]H^{(2)}_{3/2}(-k\tau)  -i\pi J_{3/2}(-k\tau) - \frac{2}{k\tau}H^{(1)}_{1/2}(-k\tau)\right)/H^{(1)}_{3/2}(-k\tau) + \frac{i\pi}{2}.}
We focus on the $\delta\log(-H\tau)$ contribution first. Note that there are six contributions from an inverse scale factor that is implicit in the canonical normalization of (\ref{drc}) and (\ref{drg}) for each mode function entering the loop integral, to be multiplied by two factors of the scale factor from the integration measures for $\tau_1$ and $\tau_2$. As argued by Senatore and Zaldarriaga \cite{SZ}, the subsequent time integrals have the form
\eq{}{\int_{-\infty}^{0}d\tau_2\int^{\tau_2}_{-\infty}d\tau_1\,\tau_1^2\tau_2^2\,e^{i\tau_2(q + \bar q - k)}e^{-i\tau_1(q + \bar q + k)}\left[...\right],}
which are dominated by contributions around $\tau_1 \sim \tau_2 \sim 1/k$ -- the time of horizon crossing of the $k$-mode -- so that the net effect after multiplying the integrands in (\ref{II}) and (\ref{I}) with factors of $\log(-H\tau_{1,2})$ and performing the time integrals will be an identical momentum integration as in (\ref{IpII}), but now with a single multiplicative factor of $\log(c' H_k\tau_k)$ where $c'$ is some order one constant. Therefore the net effect of this correction will be to correct the r.h.s. of (\ref{drint}) with the additional term
\eq{cdep}{\lim_{\delta\to 0}\delta\times\log(-cH\tau_k) k^{3 + \delta}F(\delta)}
so that (\ref{drfint}) becomes
\begin{eqnarray}
\int d^3q\,d^3\bar q\,\delta^3(\Q + \bar\Q + \K)f(q,\bar q,k) &=& k^{3}\left(F_0\, {\rm log}\,(k/\mu) + F_0 \log(-H\tau_k)+ \Lambda'\right)\\ \nn &=& k^3 \left(F_0 \log(H/\mu) + \Lambda'\right)
\end{eqnarray}
where we absorb $\log c'$ into $\Lambda$, and where we have used the horizon crossing relation $-\tau_k = 1/k$. 

Now we consider the term proportional to $u(-k\tau)$ in (\ref{wfcdr}), realizing that this term will be of the form $\delta\times k^3 (k/\mu)^{\delta}\widetilde F(\delta)$, where $\widetilde F$ is dimensionless and is the result of the momentum integrations one has to do after factoring in $u(-k\tau)$ corrections (\ref{ukt}). In order for this to contribute a logarithmic running, $\widetilde F$ is required to have a double pole in $\delta$, which can be shown not to be the case \cite{SZ}, so we are left with the expression  
\eq{dsans}{\frac{k^3}{2\pi^2}\langle \gamma^s_{ij, \K}(\tau)\gamma^{s'}_{ij,\Kp}(\tau) \rangle_{(2)} \supset \frac{N}{8\pi^2}\frac{H^4}{\mpl^4}\delta^{ss'}\delta^{3}(\K + \Kp)\left(\frac{12\pi}{5}\log(H/\mu) + \Lambda'\right) }
At this stage it might appear as if the correlation function does not run whatsoever. This cannot be the case generically, as quantum corrections typically induce running unless we are at a fixed point of the theory e.g. in the dS limit where an exact dilatation (i.e. scale) invariance is realized. Since the arguments of \cite{SZ} make it clear that corrections to the correlation functions are being forged as modes exit the horizon, \textit{it must be the case that what appears above is in fact $H_k$} -- the Hubble scale at the time the $k$-mode exits the horizon. We demonstrate this explicitly in the following by repeating the argument of \cite{SZ} including slow roll corrections to the mode functions in the loop integral.

\subsection{Dimensional Regularization on a quasi dS background}
We now retrace the steps above, but carefully considering slow roll corrections to the mode functions in the loop integral away from the dS limit, 
erring on the side of detail. To this end, we work in the limit of constant but non-zero $\epsilon := -\dot H/H^2$, since we are only interested in the leading order corrections in slow roll. We note that the defining equation for $\epsilon$ can be solved explicitly in cosmological time --
\eq{}{H(t) = \frac{1}{\epsilon t + H_*^{-1}},}
which we can integrate once more to obtain an exact expression for the scale factor
\eq{sfqds}{a(t) = a_0(1 + \epsilon H_* t)^{1/\epsilon}.}
In the above, $H_*$ is some pivot scale which we specify shortly. Note that in the limit $\epsilon \to 0$, the above is none other than the limiting expression for the usual exponential scale factor
\eq{}{\lim_{\epsilon \to 0} a(t) = a_0\, e^{H_*t}.}
Switching to conformal time, (\ref{sfqds}) becomes
\eq{atau}{a(\tau) = \frac{1}{\left(-H_*\tau\right)^{\frac{1}{1-\epsilon}}}}
where the normalization is taken to be such that the scale factor is $a = -1/(H_*\tau)$ in the dS limit. Everything so far is exact in the constant $\epsilon$ limit, but we are primarily interested in the limit $\epsilon \ll 1$, so that in what follows, we simply take
\eq{sf}{a(\tau) = \frac{1}{\left(-H_*\tau\right)^{1 + \epsilon}}}
Where from now on, we understand that we work to leading order in $\epsilon$. We now note that in order to arrive at the mode fucntions in $D = 3 + \delta$ spatial dimensions, we need to consider the action for a generic member of the $\chi$ fields in conformal time --
\eq{}{S = \frac{\mu^\delta}{2}\int d^{4 + \delta}x\,a^{2 + \delta}\left[\chi'\chi' - \partial_i\chi\partial_i\chi\right]}
making the field redefinition $v_\chi := z_\chi \chi$, where
\eq{zc}{z_\chi = a^{1 + \delta/2}\mu^{\delta/2}}
results in the canonically normalized (Mukhanov-Sasaki) action
\eq{MS}{S = \frac{1}{2}\int d^{4 + \delta}x\left[v'^2 - (\partial v)^2 + \frac{z''}{z}v^2\right].}
Similarly, the field redefinition with corresponding
\eq{zg}{z_\gamma = a^{1 + \delta/2}\mu^{\delta/2}\mpl} 
will bring the action for the individual graviton polarizations into the form (\ref{MS}) \footnote{Recalling that the $1/8$ prefactor of (\ref{s2t0}) is accounted for by the normalization of the polarization tensors (\ref{b6}).}. The equations of motion for the Fourier modes that result from (\ref{MS}) will be of the form
\eq{mseom}{v''_\K + \left(k^2 - \frac{\nu^2 - 1/4}{\tau^2}\right)v_\K = 0}
with the identification
\eq{zpp}{\frac{z''}{z} = \frac{\nu^2 - 1/4}{\tau^2} = \frac{\lambda(1 + \lambda)}{\tau^2}}
where using (\ref{sf}), we find that for both (\ref{zc}) and (\ref{zg})
\eq{}{\lambda := (1 + \epsilon)\left(1 + \delta/2\right). } 
From (\ref{zpp}) we see that $\nu = \lambda + 1/2$, which we write as
\eq{}{\nu = \frac{3 + \bar\delta}{2}}
where we have defined
\eq{bard}{\bar\delta  = \delta(1 + \epsilon) + 2\epsilon.}
Given that the mode functions that solve (\ref{mseom}) corresponding to the Bunch-Davies vacuum are given by
\eq{}{v_\K = \frac{\sqrt\pi}{2}e^{i\left(\nu + \frac{1}{2}\right)\frac{\pi}{2}}\sqrt{-\tau}H^{(1)}_\nu(-k\tau),}
one readily obtains the relevant mode functions on a $3 + \delta$ dimensional quasi de Sitter background $\chi_\K = v_\K/z_\chi$ and $\gamma_\K = v_\K/z_\gamma$ to be
\eq{drcsr}{\chi_\K(\tau) = -\frac{\sqrt{\pi}}{2}e^{i\pi\bar\delta/4}\frac{H_*^{1 + \bar\delta/2}}{\mu^{\delta/2}} \frac{(-k\tau)^{(3 + \bar\delta)/2}}{k^{(3 + \bar\delta)/2}}H^{(1)}_{(3 + \bar\delta)/2}(-k\tau),}
\eq{drgsr}{\gamma_\K(\tau) = -\frac{\sqrt{\pi}}{2}e^{i\pi\bar\delta/4}\frac{H_*^{1 + \bar\delta/2}}{\mpl\,\mu^{\delta/2}} \frac{(-k\tau)^{(3 + \bar\delta)/2}}{k^{(3 + \bar\delta)/2}}H^{(1)}_{(3 + \bar\delta)/2}(-k\tau).}
which are identical to (\ref{drc}) and (\ref{drg}) with the replacement $\delta \to \bar\delta$ everywhere except in the power of $\mu^\delta$ in the denominator, which book keeps the mass dimension of the Fourier component fields. As before, we expand in powers of $\bar\delta$ to obtain
\eq{wfcdrsr}{\chi_{\K}(\tau) = \frac{iH_*}{\sqrt{2\mu^\delta k^3}}\left(1 + i k\tau\right)e^{-i k\tau}\left[1 + \frac{\bar\delta}{2}\log(-H_*\tau) + \frac{{\bar\delta}^2}{8}\log^2(-H_*\tau) + 
	\frac{\bar\delta}{2} u(-k\tau) + \frac{{\bar\delta}^2}{8} v(-k\tau) +...\right] }
with a similar expansion for $\gamma_{\K}(\tau)$. The function $u(-k\tau)$ is given by (\ref{ukt}), now with $\bar\delta$ given by (\ref{bard}), and $v(-k\tau)$ is a term that arises from the second derivative of the Hankel function with respect to its argument, whose explicit form will not be necessary in what follows. The reason we need to go to second order in $\bar \delta$ can be anticipated from the fact that even though we neglect terms of order $\delta^2$ and $\epsilon^2$, we still have $\bar\delta^2 = 4\epsilon\delta$, which can contribute leading slow roll corrections to finite terms when multiplying $\delta$ poles.

\subsubsection{Slow-roll corrected loop integral}
The net result of inserting the slow roll corrected mode functions and scale factor in the dimensionally regularized integral (\ref{drint}) to second order in $\bar{\delta}$, is to effect the corrections
\begin{align}
\label{src}
\int d^3q\,d^3\bar q\,\delta^3(\Q + \bar\Q + \K)f(q,\bar q,k)\\&&\hspace{-80pt}\nn = k^3\left(\frac{k}{\mu}\right)^\delta \Biggl[\left(\frac{F_0}{\delta} + F_1\right)\left(1 + \bar{\delta}\log(- H_*\tau_k) + \frac{\bar{\delta}^2}{2}\log^2(- H_*\tau_k)\right) + 3\bar{\delta}\left(\frac{\bar F_0}{\delta} + \bar F_1\right)\Biggr] \label{src}
\end{align}
The single log term is the result of multiplicative corrections of the form $\frac{1}{2}\bar{\delta}\log[-H_*\tau_k]$ from each of the six momentum independent corrections (cf. footnote \ref{footnote}) to the mode functions that run through the loops in (\ref{wfcdrsr}), compensated by the single log contributions from the dimensionally regularized measures of each contributing interaction Hamiltonian:
\eq{meassr}{a^{2(2 + \delta)} = (-H_*\tau)^{-2(1 + \epsilon)(2 + \delta)} = \frac{1}{(-H_*\tau_k)^4}\left(1 -2 \bar{\delta}\log(- H_*\tau_k) + 2\bar{\delta}^2\log^2(- H_*\tau_k)\right) }
Similarly, one can collect all double log terms from the mode functions and the measure to result in the contribution proportional to $\frac{\bar{\delta}^2}{2} = 2\epsilon\delta$ above\footnote{This comes from collecting all terms quadratic in $\bar{\delta}\log{(-H_*\tau_k)}$ from the product of six factors of the term in the square brackets of (\ref{wfcdrsr}) and the logarithmic contributions from the measures (\ref{meassr}).}. As argued in the previous subsection, all momentum independent corrections serve to multiply the loop integral resulting from inserting the uncorrected mode functions, but with log factors resulting from the fact that these integrals are dominated by contributions at horizon crossing. The term proportional to $3\bar{\delta}$ in the above is the leading order contribution from the loop integrations incorporating the momentum dependent corrections to the mode functions, given by $u(-k\tau)$ and $v(- k\tau)$ in (\ref{wfcdrsr}), which we will not need to calculate explicitly in what follows
. In spite of appearances, the last term of (\ref{src}) does indeed factor in momentum dependent corrections to the wavefunctions of order $\bar{\delta}^2 = 4\epsilon\delta$, whose effects can simply be absorbed into the definition of $\bar F_0$ and $\bar F_1$.

One immediately sees that slow roll corrections to the loop integrals result in additional $\delta$ poles that go as $\epsilon/\delta$, which must correspond to slow roll corrections to the counter terms. We will now show that this is explicitly the case by calculating these counterterms. The final outcome will be that the factor in (\ref{dsans}) becomes
\eq{}{\log(H_*/\mu) \to \log(H_k/\mu)}
where $H_k$ is the Hubble factor at horizon crossing of the comoving scale $k$. Let us now show this explicitly. We first observe that expanding (\ref{src}), collecting terms, and recalling that $\bar{\delta} = \delta(1 + \epsilon) + 2\epsilon$ and $\bar{\delta}^2 = 4\epsilon\delta$, the net result of dimensionally regularizing the loop integral is the correction
\begin{eqnarray}
\label{qdsans}\nn \frac{k^3}{2\pi^2}\langle \gamma^s_{ij, \K}(\tau)\gamma^{s'}_{ij,\Kp}(\tau) \rangle_{(2)} &=& \frac{N}{8\pi^2}\frac{H_*^4}{\mpl^4}\delta^{ss'}\delta^{3}(\K + \Kp)\Biggl\{\highlight{F_0\left[\log \frac{k}{\mu} + (1 + \epsilon)\log\frac{H_*}{k} \right]} + \frac{F_0}{\delta}\left[1 + 2\epsilon\log\frac{H_*}{k}\right]\\ \nn  &&\hspace{-120pt}+ F_1\left[1 + 2\epsilon\log\frac{H_*}{k}\right] + 2\epsilon F_0\log^2\frac{H_*}{k}+ 2\epsilon F_0\log\frac{k}{\mu}\log\frac{H_*}{k} + 6\epsilon \bar F_0\log\frac{k}{\mu} + 6\epsilon\frac{\bar F_0}{\delta}  + 3(1+\epsilon)\bar F_0 + 6\epsilon\bar F_1\Biggr\}\\ \label{expr}
\end{eqnarray}
where the highlighted term in the above is only contribution that will survive after we subtract the divergences and imposing the renormalization conditions at some finite scale. As is standard in effective field theory, one is entitled to pick the renormalization scale to minimize the logarithms at any particular scale of interest (typically representing some mass threshold), which in the present context is naturally given by $\mu = H_*$. Doing so results in the cancellation of the double log contributions, so that
\begin{eqnarray}
\label{qdsans}\nn \frac{k^3}{2\pi^2}\langle \gamma^s_{ij, \K}(\tau)\gamma^{s'}_{ij,\Kp}(\tau) \rangle_{(2)} &=& \frac{N}{8\pi^2}\frac{H_*^4}{\mpl^4}\delta^{ss'}\delta^{3}(\K + \Kp)\Biggl\{\highlight{F_0\left[\log \frac{k}{\mu} + (1 + \epsilon)\log\frac{H_*}{k} \right]} + \frac{F_0}{\delta}\left[1 + 2\epsilon\log\frac{H_*}{k}\right]\\  &+& F_1\left[1 + 2\epsilon\log\frac{H_*}{k}\right] + 6\epsilon \bar F_0\log\frac{k}{\mu} + 6\epsilon\frac{\bar F_0}{\delta}  + 3(1+\epsilon)\bar F_0 + 6\epsilon\bar F_1\Biggr\} \label{expr2}
\end{eqnarray}
We first focus on the highlighted term to show that indeed
\eq{hk0}{\highlight{\log \frac{k}{\mu} + (1 + \epsilon)\log\frac{H_*}{k}} = \log\frac{H_k}{\mu}}
From (\ref{sf}) we see that $H = \dot a/a = a'/a^2 = H_*(1 + \epsilon)(-H_*\tau)^\epsilon$. Furthermore, the horizon crossing condition $k = aH$ implies that the $k$-mode exits the Hubble radius at $\tau_k = -\frac{(1 + \epsilon)}{k}$, so that 
\eq{hk}{H_k := H(\tau_k) = H_*(1+\epsilon)\left(\frac{H_*(1+\epsilon)}{k}\right)^\epsilon = k\left(\frac{H_*(1+\epsilon)}{k}\right)^{1+\epsilon} = k\left(\frac{H_*}{k}\right)^{1+\epsilon},}
where the difference between the second and third terms is higher order in $\epsilon$.
Hence $H_k/k = (H_*/k)^{1+\epsilon}$ and the final equality in (\ref{hk0}) immediately follows. It now remains to show that all the un-highlighted terms in (\ref{expr2}) will be subtracted by the relevant slow-roll corrections to the counterterms.

\subsubsection{Slow-roll corrected counterterms}
We are calculating in an effective theory with a cut-off scale $\Lambda$. As discussed in the following section of the appendix, since we are considering only the spin two perturbations generated from the Einstein-Hilbert term in the presence of $N$ scalar fields, this cut-off is given by
\eq{}{\Lambda \sim \frac{4\pi\mpl}{\sqrt{N}}}
Where we absorb any order unity coefficients of the precise cutoff in the coefficients of the counterterms we are about to calculate. From the structure of the divergences, we see that these counterterms will take the form of all possible dimension six operators consistent with the symmetries of the problem. Hence
\eq{counter}{H_{c.t.} = \mu^\delta N \int d^{3+\delta}x\, a^\delta\left[c_1\partial_k\gamma'_{ij}\partial_k\gamma'_{ij} + c_2\partial_l\partial_k\gamma_{ij}\partial_l\partial_k\gamma_{ij} + c_3\gamma''_{ij}\gamma''_{ij}\right]} 
where we note that the pre-factor of $\mpl^2$ has been canceled in dividing by $\Lambda^2$. We also note a non-standard feature of the effective theory of inflation -- although the coefficients of the dimension six operators are constants (i.e. Wilson co-efficients), one could in principle add counterterms where the coefficients are functions of time (i.e. \textit{Wilson functions}), and appear at dimension six as operators of the form 
\eq{}{\Delta H_{c.t.} =  \mu^\delta N \int d^{3+\delta}x\, a^\delta\left[ c_4(a H)^2\gamma'_{ij}\gamma'_{ij} + c_5(a H)^2\partial_k\gamma_{ij}\partial_k\gamma_{ij}\right]} 
with $c_4, c_5$ dimensionless. That these must appear can be seen from the fact that field redefinitions using the unperturbed equations of motion will necessarily generate such terms. Although they will not be necessary for the renormalization we are about to perform, loop corrections to more complicated processes will necessitate them, particularly if we are interested in calculating corrections to correlation functions at finite times (and not $\tau \to 0$). 

Focusing on the counterterms (\ref{counter}), we need to calculate
\begin{eqnarray}
\langle \gamma^s_{ij, \K}(\tau)\gamma^{s'}_{ij,\Kp}(\tau) \rangle_{c.t.} &=& i\int_{-\infty_-}^\tau\langle[H_{c.t.}(\tau'),\gamma^{0,s}_{ij, \K}(\tau)\gamma^{0,s'}_{ij,\Kp}(\tau) ] \rangle d\tau'\\ \nn &=& 2\Im\,\int_{-\infty_-}^\tau \langle\gamma^{0,s}_{ij, \K}(\tau)\gamma^{0,s'}_{ij,\Kp}(\tau)H_{c.t.}(\tau') \rangle d\tau'\\ \nn &=& 64N(2\pi)^3\mu^\delta\delta^{ss'}\delta^{3}(\K + \Kp)\gamma_k^2(\tau)\\ \nn &\times& \Im\int_{-\infty_-}^\tau d\tau' a^\delta \left\{c_1 k^2 \gamma_k'^{*2}(\tau') + c_2 k^4 \gamma_k^{*2}(\tau') + c_3 \gamma_k''^{*2}(\tau') \right\} 
\end{eqnarray}
It is straightforward to calculate the above with the first order corrections to the dimensionally deformed, slow roll corrected mode functions (\ref{wfcdrsr}) using (\ref{ukt}), including slow roll corrections to the dimensionally deformed measure $a^\delta$. The resulting integrals can be performed analytically in the $\tau \to 0$ limit\footnote{\label{footnote} We note that we didn't need to dimensionally regularize the mode functions that arose from the external legs in the loop integral (\ref{IpII}) since these will be compensated by identical terms from the external legs in the counterterms \cite{SZ}. The net result is equivalent to only dimensionally regularizing terms that depend on loop momenta. Similarly, we don't need to consider slow roll corrections to the external wavefunctions in the above since these will simply add a tilt to the overall spectral index multiplying both the loop integrals and the counterterm contributions.}. The counterterm contributions to the regularized dimensionless power spectrum are given by
\begin{eqnarray}
\nn \frac{k^{3}}{2\pi^2}\langle \gamma^s_{ij, \K}(\tau)\gamma^{s'}_{ij,\Kp}(\tau) \rangle_{c.t.} &=& 16\pi N \delta^{ss'}\delta^{3}(\K + \Kp)\frac{H_*^4}{\mpl^4}\Bigg\{ \epsilon\left(-c_1 + c_2 + 5 c_3\right) -6 c_2(1+\epsilon)\delta \\ &+& \left(c_1 - 5c_2 - c_3\right)\left[1 + 2\epsilon\left(\log\frac{H_*}{k} -\gamma_E - \log 2 + 2\right)\right]\Biggr\}
\end{eqnarray}
We note that the terms proportional to $2\epsilon$ in the second line above are simply the first order slow roll correction to the square of the wavefunction modulus at long wavelengths. Furthermore, in the combination $1 + 2\epsilon \log(H_*/k)$, we see precisely the slow roll correction to the counterterm needed to cancel the single $\delta$ pole encountered in the dS case. However, other $\delta$ poles now appear in (\ref{qdsans}), in addition to to finite terms that must be fixed by renormalizing at a particular scale. It is straightforward to check that the choices
\begin{eqnarray}
128\pi^3 c_2 &=& \frac{1}{\delta}\left[\frac{1}{2}\left(2-\epsilon\right)\bar F_0 + \epsilon \bar F_1\right]\\
128\pi^3(-c_1 + 5c_2 + c_3) &=& \frac{F_0 + 6\epsilon \bar F_0}{\delta} + F_1 - 3\bar F_0\\
128\pi^3(c_1 - c_2 - 5 c_3) &=& \left(6\bar F_0-2\frac{F_0}{\delta} - 2F_1\right)\left[2 - \gamma_E - \log 2\right] 
\end{eqnarray}
cancel all unhighlighted terms in (\ref{qdsans}), thus fixing the renormalization condition at $\mu = H_*$, which we take to be the pivot scale at which we fix the tensor to scalar ratio (i.e. the loop correction is normalized to vanish there). Hence 
\eq{}{\frac{k^3}{2\pi^2}\langle \gamma^s_{ij, \K}(\tau)\gamma^{s'}_{ij,\Kp}(\tau) \rangle_{(2)} = \frac{N}{8\pi^2}\frac{H_*^4}{\mpl^4}\delta^{ss'}\delta^{3}(\K + \Kp)F_0 \log \frac{H_k}{H_*} }
with $F_0 = 12\pi/5$ as calculated in the previous subsection. 

We note that the net effect of including slow roll corrections is to reintroduce a $\log k/k_*$ running to the loop correction, but with extra $\epsilon$ suppression and with the opposite sign. This follows from the fact that $\log \frac{H_k}{H_*} = - \epsilon_* \log \frac{k}{k_*} + \mathcal O(\epsilon_*^2)$, which we see as follows. Consider the expression (\ref{sf}) as well as $H=a'/a^2$ where prime  denote derivatives w.r.t. $\tau$.
This gives $H= (1+\epsilon)H_*(-H_*\tau)^\epsilon$. With $-\tau_k=k$, recalling that we have identified $H_*$ as the pivot scale at which we measure the tensor to scalar ratio, and $\epsilon=\epsilon_*$, we obtain
\eq{htok}{\log \frac{H_k}{H_*} = - \epsilon_* \log \frac{k}{k_*} + \mathcal O(\epsilon_*^2) \,.
}
so that 
\eq{sr2cf}{\frac{k^3}{2\pi^2}\langle \gamma^s_{ij, \K}(\tau)\gamma^{s'}_{ij,\Kp}(\tau) \rangle_{(2)} = -\epsilon_*\frac{N}{8\pi^2}\frac{H_*^4}{\mpl^4}\delta^{ss'}\delta^{3}(\K + \Kp)F_0 \log \frac{k}{k_*}.}
Comparison with (\ref{pstdef}) after summing over polarizations implies the loop correction to the tensor power spectrum
\eq{sr2cf2}{\Delta\calP_{\rm \gamma, 1-loop} = - \epsilon_*\frac{2 H_*^2}{\pi^2\mpl^2}\frac{N}{16\pi^2}\frac{H_*^2}{\mpl^2}\frac{3}{5} \log \frac{k}{k_*}.}

\section{The strong coupling bound for gravity}
\label{a:sc}
We first present an effective field theory derivation of the strong coupling bound for gravity (as discussed in \cite{Dvali:2007hz, Dvali:2007wp}), namely that in the presence of $N$ species, one can only treat gravity semi-classically for scales less than
\eq{scb}{\Lambda \lesssim \frac{4\pi\mpl}{\sqrt N}.}
This is easily seen from the fact that were we only interested in calculating n-point correlation functions of gravitons, then the relevant quantities can be reproduced from the effective action where the matter fields have been integrated out \cite{BD, HK}
\eq{}{S = S_{\rm EH} + \Delta W_2 + ...}
where $S_{\rm EH}$ is the usual Einstein-Hilbert action, and where the leading quadratic curvature corrections have the form
\begin{equation}
\begin{split}
\label{hkea}
\triangle W_2 = \frac{1}{2880 \pi^2} \int \sqrt{-g} \Bigg[ (b+2a) R_{\mu \nu} R^{\mu \nu}+ \Big( c - \frac{b + 2a}{3} \Big) R^2 \Bigg]
\end{split}
\end{equation}
where the coefficients $a,b,c$ depend on the spin (and in the case of scalars, the non-minimal coupling parameter $\xi$) of the field integrated out \cite{HK}:
\begin{center}
	\begin{tabular}{p{1cm} | p{1cm} | p{1cm} | l }
		spin & a & b &  c\\
		\hline
		0&1&1& 90$(\xi- 1/6)^2$\\
		1/2 & $-7/2$ & $-11$& 0\\
		1 & $-13$ & 62& 0\\
		2 & 212 & 0 &717/4\\
	\end{tabular}
\end{center}
Note that in the above, we have eliminated redundant operators by using the Gauss-Bonnet identity in 4D. For $N$ species of various spins, the coefficients that sit in front of the curvature squared terms will be a weighted sum of the above. From this, one can see that the contribution from the quadratic terms in the effective action become comparable to those from the Einstein-Hilbert term at momentum transfers approaching $p^2 \sim \kappa \mpl^2/N$, or for background curvatures approaching
\eq{scbeft}{R \sim \kappa \frac{\mpl^2}{N}} 
where $\kappa$ is a spin weighted sum, and $N$ is the number of species we have integrated out. For $N$ minimally coupled scalar fields, we find from (\ref{hkea}) and the table above that on any maximally symmetric background, $\kappa = 40\cdot 16\pi^2$. On a dS background, $R = 12 H^2$ so that the implication of the strong coupling bound is that
\eq{}{H^2_{\rm dS} \lesssim \frac{10}{3}\frac{16\pi^2\mpl^2}{N},}
or that the effective theory is only reliable for momentum transfers up to the scale $\Lambda \lesssim 4\pi\mpl/\sqrt N$, where we have neglected an order unity spin dependent coefficient. 

Another argument presented in \cite{Dvali:2007wp} concerns black hole evaporation, where it was argued that black holes of the size $\Lambda^{-1}$ also have a lifetime of $\Lambda^{-1}$, suggesting that this is the scale at which quantum gravity becomes relevant. Consider the evaporation rate for a black hole of mass $M$ \cite{Carlip:2014pma}:
\eq{}{\frac{d M}{dt} = -\varepsilon\sigma T_H^4 A}
where $\sigma$ is the Stefan-Boltzmann constant, $T_H$ is the Hawking temperature of the black hole and $\varepsilon$ is the grey body emmissivity factor that is proportional to the number of species the black hole can radiate: 
\eq{}{\varepsilon \propto N.}
Denoting the constant of proportionality $c_\varepsilon$ so that $\varepsilon = c_\varepsilon N$, we then have
\eq{life}{\frac{d M}{dt} = -Nc_\varepsilon\sigma \frac{\mpl^4}{4\pi M^2}}
where we have used the fact that the Schwarzschild radius of a black hole of mass $M$ is given by $R_s = M/(4\pi\mpl^2)$ and $T_H = \mpl^2/M$ in reduced Planck units, so that (\ref{life}) can be expressed as
\eq{life2}{\tau_{\rm BH} = \frac{4\pi}{N}\int_0^{M_{\rm in}} \frac{M^2 dM}{c_\varepsilon\sigma \mpl^4} = \frac{4\pi}{N} \frac{M_{\rm in}^3}{3 c_\varepsilon\sigma \mpl^4},}
where $M_{\rm in}$ is the initial size of the black hole. Now consider a black hole of initial radius 
\eq{qgd}{R_s = \frac{M_{\rm in}}{4\pi\mpl^2} := \Lambda^{-1}_{\rm QG},}
where $\Lambda_{\rm QG}$ is understood to be defined by the above. Setting the lifetime to equal the size of the black hole allows us to determine the scale at which quantum gravity must become relevant. Setting (\ref{life2}) equal to (\ref{qgd}), and using the latter to express $M_{\rm in}$ in terms of $\Lambda_{\rm QG}$, we see that this is when
\eq{}{\Lambda_{\rm QG} \sim \lambda \frac{4\pi\mpl}{\sqrt{N}}}
with $\lambda = 4\pi/(3\sigma c_\varepsilon)$. Given that $\sigma = \pi^2k_B^4/(60\hbar^3 c^2)$ $\sim 0.5$ in natural units, and given that the emissivity constant $c_\varepsilon$ will necessarily less than unity for a grey body, we see that $\lambda > 1$ and the strong coupling bound is (\ref{scb}) is actually stronger than the one derived here. 
\end{appendix}

\end{document}